\documentclass[a4paper]{article}

\usepackage[normalem]{ulem}
\usepackage{titlesec}
\usepackage{comment}
\usepackage[english]{babel}
\usepackage[T1]{fontenc}
\usepackage{amssymb}
\usepackage{amsthm}
\usepackage[a4paper,top=2.5cm,bottom=2.5cm,left=2cm,right=2cm,marginparwidth=1.75cm]{geometry}
\usepackage{enumerate}
\usepackage{wrapfig}
\usepackage{enumerate}
\usepackage{enumitem}

\usepackage[dvipsnames]{xcolor}
\usepackage{lineno}

\usepackage{verbdef}
\verbdef{\vtext}{verb text}

\newcommand{\var}{\operatorname{Var}}
\newcommand{\es}{e_{\textbf{S}}}
\newcommand{\diag}{\operatorname{diag}}

\newcommand{\T}{\operatorname{\,T}}

\newcommand{\logit}{\operatorname{logit}}

\usepackage[export]{adjustbox}
\usepackage{subfig}
\usepackage{amsmath}
\usepackage{graphicx}
\usepackage[colorinlistoftodos]{todonotes}
\usepackage[colorlinks=true, allcolors=blue,hypertexnames=false]{hyperref}
\usepackage{float}
\usepackage{authblk}

\usepackage{amsmath,amsfonts,amssymb,amsthm}
\usepackage{mathtools}
\usepackage{commath}

\title{Syphilis core groups informed by network connectivity and treatment} 
\title{{Pre-exposure prophylaxis and syphilis in men who have sex with men: a network analysis}}
\author[1]{Esteban Vargas Bernal}
\author[2]{Morgan Spahnie}
\author[3]{William C. Miller}
\author[2,4]{Abigail N. Turner}
\author[2,5]{Joseph H. Tien}
\affil[1]{School of Mathematical and Statistical Sciences, Arizona State University}
\affil[2]{College of Public Health, The Ohio State University}
\affil[3]{Gillings School of Global Public Health, University of North Carolina}
\affil[4]{College of Medicine, The Ohio State University}
\affil[5]{Department of Mathematics, The Ohio State University}

\newtheorem{definition}{Definition}

\newtheorem*{proposition*}{Proposition}

\usepackage{algorithmic}            
\usepackage{algorithm}

\titleformat{\paragraph}
{\normalfont\normalsize\bfseries}{\theparagraph}{1em}{}
\titlespacing*{\paragraph}
{0pt}{3.25ex plus 1ex minus .2ex}{1.5ex plus .2ex}

\newcounter{savealgorithm}

\begin{document}
\maketitle

\begin{abstract}

Pre-exposure prophylaxis (PrEP) has been established as an effective tool for preventing HIV infection among men who have sex with men (MSM).  However, there is the possibility of PrEP usage leading to increased sexual partners and increased transmission of non-HIV sexually transmitted infections such as syphilis.  We take here a network perspective to examine this possibility using data on sexual partnerships, demographic data, PrEP usage, and syphilis among MSM in Columbus, Ohio.  We use a recently developed community detection algorithm, an adaptation of the {community detection} algorithm InfoMap to absorbing random walks, to identify clusters of people (`communities') that may drive syphilis transmission.  Our {community detection} approach takes into account both sexual partnerships as well as syphilis treatment rates when detecting communities.  We apply this algorithm to sexual networks fitted to empirical data from the Network Epidemiology of Syphilis Transmission (NEST) study in Columbus, Ohio. We assume that PrEP usage is associated with 
regular visits to a sexual health provider, and thus is correlated with syphilis detection and treatment rates. {We examine how PrEP usage can affect community structure in the sexual networks fitted to the NEST data.}
We identify two types of PrEP users, those belonging to a large, highly connected community and tending to have a large number of sexual partners, versus those with a small number of sexual partners and belonging to smaller communities. A stochastic syphilis model indicates that PrEP users in the large community may play an important role in sustaining syphilis transmission.

\


\end{abstract}

\noindent {\bf Acknowledgements: } 
This work was supported by the National Science Foundation (DMS-1814737) and the Centers for Disease Control (U01PS005170).

\pagebreak

\section{Introduction}\label{sec:intro}
A resurgence in syphilis cases was reported in the United States from 2013-2017, particularly among men who have sex with men (MSM), and syphilis continues to be a significant public health concern \cite{schmidt2019resurgence}. Various reasons have been suggested for this resurgence, including {geospatial networking applications (e.g. dating / hookup apps) \cite{goedel2015,holloway2015network,macapagal2018}}, illicit drug usage \cite{chesney1998histories}, and changes in attitudes towards HIV infection and associated increased sexual partners and condomless sex \cite{chen2002continuing}.  A high proportion of undetected early syphilis, a subset of people with a large number of concurrent sexual partners, and an increase in condomless sex due to HIV risk compensation \cite{kojima2016pre} are some of the challenges involved in controlling syphilis transmission.

Here we use tools from network science as a lens to examine drivers of syphilis transmission among MSM using empirical data {from Columbus, Ohio collected through the} 
Network Epidemiology of Syphilis Transmission (NEST) study \cite{copen2022factors}.  
We are particularly interested in the potential impact of pre-exposure prophylaxis (PrEP) usage for HIV on syphilis core groups and transmission.  PrEP is an established, effective tool for decreasing HIV incidence \cite{grant2010preexposure}.  PrEP usage is also associated with an increase in detection and treatment of syphilis among MSM \cite{chow2017increased,tang2020quarterly} due to increased screening for sexually transmitted infections (STIs). However, increased sexual partners and condomless sex among PrEP users may lead to increased incidence of STIs other than HIV. For example, MSM who use PrEP are more likely to be infected with syphilis than MSM who do not use PrEP \cite{kojima2016pre}. In addition, groups of PrEP users have a relatively large connectivity \cite{kuhns2017use} and might constitute core groups \cite{van2021p316}.  PrEP usage thus may simultaneously be associated with a decreased infectious period for syphilis and an increase in the number of sexual partners. It is not obvious what the net impact of PrEP usage is for syphilis transmission among MSM.

In this paper we examine how PrEP 
usage for HIV potentially shapes the effective structure of MSM sexual networks, which in turn can impact syphilis transmission. Specifically, we use community detection algorithms from network science to find highly interconnected clusters of people {with low connectivity between clusters. We call these clusters `communities' \cite{newman2004detecting}.} Importantly, our community detection approach takes into account both sexual partnerships as well as heterogeneous disease treatment rates between people.  Different treatment rates between people can impact disease propagation on networks and induce effective communities that differ from those obtained from considering only the network edges \cite{vargas2022infomap}. The NEST study provides empirical data on sexual partnerships and {demographic features (such as age and race)} among MSM, from which we derive sexual partnership networks. The NEST data furthermore provide information on PrEP usage.  We assume that PrEP usage is associated with being in contact with a sexual health provider, and is thus associated with increased rates of syphilis detection and treatment compared with people not using PrEP. The sexual networks together with treatment rates based on PrEP usage are what we use as inputs to our community {detection algorithm}.

Communities can vary in their size, composition, and connectivity. This community structure can have a marked effect on disease dynamics, for example affecting disease quantities such as outbreak size and duration \cite{salathe2010dynamics}.  An example of community detection in the context of MSM and sexually transmitted infections (STIs) is the work of Billock et al \cite{billock2020network} that includes a multi-layer analysis of HIV and syphilis networks for MSM in North Carolina. {Performing community detection on sexual networks} can identify clusters of people driving transmission, which is closely related to the concept of core groups in STI epidemiology 
\cite{cooke1973some}. Many different approaches have been used for defining and identifying core groups, including network \cite{wang2020preventing}, clinical \cite{potterat1985gonorrhea, watt1958gonorrhoea}, mathematical \cite{cooke1973some,may1987commentary}, and sociological approaches \cite{thomas1996development}.  A large number of reinfections in a subset of people is an example of a clinical criterion for a core group \cite{watt1958gonorrhoea}. Additionally, a cluster of people with a large number of sexual partners leading to a large number of secondary infections is an instance of a mathematical {criterion} for a core group \cite{cooke1973some, may1987commentary}.  

We examine the impact of PrEP usage and differential syphilis treatment rates on community structure by fitting an exponential random graph model (ERGM) to the NEST data.  ERGMs are a class of random graphs that are widely used to examine how network structure depends upon covariates of interest such as race, age, and other demographic and behavioral features \cite{krivitsky2017inference}.  Here we include PrEP usage as one of the covariates in the ERGM.  We fit the ERGM to the NEST data and use the resulting fitted ERGM to generate sexual networks for MSM in Columbus that are consistent with the empirical data.  We then perform community detection on the resulting networks, using a community detection algorithm \cite{vargas2022infomap} that adapts the widely-used algorithm InfoMap \cite{rosvall2009map,rosvall2008maps} to absorbing random walks. We refer to this community detection algorithm as InfoMap for absorbing random walks.  InfoMap for absorbing random walks
takes disease treatment rates into account \cite{vargas2022infomap}, allowing us to evaluate how PrEP usage shapes effective community structure. 

We use a stochastic syphilis model on networks to evaluate how the detected community structure impacts disease dynamics.  Different syphilis models have been developed and parameterized, including differential equation models \cite{tuite2016go} and agent-based models \cite{tuite2013screen}.  Here we adapt the compartmental model of Tuite et al \cite{tuite2016go} to a network setting.  Specifically, we consider a network where the possible states of each node match the compartments of Tuite et al's model \cite{tuite2016go}.    A stochastic model allows us to examine the distribution of quantities such as the number of infections over a specified time period, and in particular how these distributions vary between different clusters in the network.

The remainder of this paper is structured as follows. In Section \ref{sec:the_network}, we present the NEST data and the networks sampled from an ERGM fitted to the Columbus NEST data. In Section \ref{sec:syphilis_model_3}, we present the stochastic syphilis model that we use in our analysis of syphilis dynamics. In Section \ref{sec:community}, we present the communities {detected} by InfoMap for absorbing random walks. In Section \ref{sec:dynamics}, we present our results on syphilis dynamics on the communities yielded by InfoMap for absorbing random walks. In Section \ref{sec:discussion_3}, we discuss the results of Sections \ref{sec:community} and \ref{sec:dynamics}. In Appendix \ref{sec:appendix}, we present background material on ERGMs, details of the fitted ERGM, and background material on InfoMap for absorbing random walks.

\section{The networks} \label{sec:the_network}

\subsection{Columbus NEST empirical data} \label{sec:data_NEST}
%
The empirical data for this study come from the Network Epidemiology Study of Syphilis Transmission (NEST) project, a multi-site prospective cohort study examining risk factors for syphilis among MSM in Baltimore, Chicago, and Columbus.  A total of 748 participants enrolled in the NEST study \cite{copen2022factors}.  We focus here on the 241 study participants from the Columbus site {enrolled during 2019 and 2020}.  MSM at each site were recruited through a combination of approaches including convenience sampling, respondent driven sampling (RDS), and venue-based recruitment.  Sociodemographic data including age, race and ethnicity, education, and health insurance status were collected from participants at intake.  During quarterly visits, each participant listed sexual partners in the last three months, together with demographic data (such as age and race) of their partners.  Participants also completed a behavioral questionnaire during each quarterly visit.  This questionnaire included behaviors such as PrEP usage, exchange sex, group sex, and drug or alcohol use during sex in the last three months.  The NEST project is an egocentric study (egos = study participants, alters = sexual partners of study participants).  In particular, the NEST project has a minimal egocentric network study design, meaning that the alters cannot be uniquely identified for different egos and partnerships among alters are not identified either.   Further information on the NEST project design and data collection process are given in \cite{copen2022factors}.

We represent the NEST network as a sub-network of the MSM sexual network in Columbus, where the set of egos is denoted by $\mathbf{S}$, and the set of alters that are reported by an ego $i$ are the neighbors of $i$ in the sub-network. Let $e_i^{\textnormal{e}}$ denote the attribute values of ego $i$, and let $e_i^{\textnormal{a}}$ denote the unordered list of the attribute values of the alters of ego $i$. Let $\es$ be the set of unordered pairs $\{e_i^{\textnormal{e}}, e_i^{\textnormal{a}}\}$. The set $\es$ is called {an} \emph{egocentric sample} of the MSM sexual network (see Appendix \ref{sec:egocentric_samples} for more details on egocentric samples).  The node attribute values that we consider are listed in the second column of Table \ref{tab:descriptive_ergm_data}. More specifically, we use age, broken into four age groups (age group 1: $[18, 25)$, age group 2: $[25,35)$, age group 3: $[35,45)$, age group 4: 45+), race (\emph{white, black, other}), PrEP usage in the last three months prior to at least one visit (\emph{yes, no}), sex trade in the last three months prior to at least one visit (\emph{yes, no}), and group sex in the last three months prior to at least of one visit (\emph{yes, no}). From the 241 egos in the NEST network, there were five egos whose alters had missing attributes among age and race. These five egos and the corresponding alters are removed from the sample to obtain an egocentric sample $e_\mathbf{S}$ with $|\mathbf{S}| = 236$ egos and 2461 alters.  

We refer to a node that corresponds to a person who has used PrEP  as a \emph{PrEP node}, and use {\it PrEP status} to refer to whether a node is either a PrEP node or a non-PrEP node. In the third column of Table \ref{tab:descriptive_ergm_data}, we give the distributions of the attributes of $e_{\mathbf{S}}$. {We observe that the egocentric sample $e_{\mathbf{S}}$ has 77 PrEP nodes (32\% of the 236 egos in the data).} The mean degree of the egos in $e_{\mathbf{S}}$ is 10.43 $(sd = 11.82)$. The conditional mean degree for PrEP egos of $e_{\mathbf{S}}$ is 15.1 $(sd = 13.9)$ and the conditional mean degree for non-PrEP egos of $e_{\mathbf{S}}$ is 8.18 $(sd = 9.98)$. In Figure \ref{fig:degree_by_tr}(a), we show the conditional degree distributions of $e_{\mathbf{S}}$. We observe that the degree distribution conditioned to PrEP nodes is heavier to the right than the corresponding degree distribution conditioned to non-PrEP nodes. We refer to the proportion of nodes with degree one as the \emph{monogamy proportion}. The monogamy proportion in $e_{\mathbf{S}}$ is 10.59\%.

\subsection{Exponential random graph models}\label{sec:fitted_to_NEST}

We use a family of exponential random graphs to simulate sexual networks based on the NEST data on sexual partnerships and demography.  Exponential random graph models (ERGMs) are a widely used family of random graph models, and robust statistical inference approaches have been developed for fitting ERGMs to egocentric data \cite{krivitsky2017inference}.  We give here a brief description of ERGMs for the convenience of the reader.  Further discussion of ERGMs can be found in Appendices \ref{sec:ERGMs_basics} and \ref{sec:ERGM_statistics} and in \cite{krivitsky2017inference}.  

A linear exponential random graph model (linear ERGM) $\mathbb{P}_\theta$ is a distribution on the space $\mathcal{Y}^{(n)}$ of undirected graphs of size $n$ \cite{hunter2006inference,krivitsky2017inference}. Let $Y$ be a random variable on the space $\mathcal{Y}^{(n)}$.  Then the distribution $\mathbb{P}_\theta$ is defined by 
\begin{align}
    \mathbb{P}_{{\theta}}(Y=y) = \frac{\exp\left({\theta}^{\T}g(y)\right)} {\sum_{z\in \mathcal{Y}^{(n)}} \exp\left(\theta^{\T} g(z)\right)} \,,
\label{eqn:P_ergm}
\end{align}
where $\theta =\left(\theta_1,\ldots,\theta_{k}\right)^{\T}$ is a coefficient vector and $g(y) = (g_1(y), \ldots, g_{k}(y))^{\T}$ is a vector of statistics that depend upon the edges and node attributes of $y$.  The statistics $g_i(y)$ are quantities that potentially influence network structure.  For example, sexual networks may exhibit homophily with respect to race or other demographic quantities.  
We may then include a homophily statistic $g_A(y)$ in \eqref{eqn:P_ergm} that is equal to the number of edges of $y$ with both nodes in some set $A$ (e.g., $A$ corresponding to nodes belonging to a particular racial group).

The coefficient vector $\theta$ quantifies how network structure depends upon the statistics in $g$.  In particular, the conditional probability of an edge connecting nodes $i$ and $j$ can be expressed in terms of $\theta$ and the change in the statistics in $g$ resulting from toggling an edge between $i$ and $j$ from off to on while keeping the rest of the network fixed.  Given a network $y \in \mathcal{Y}^{(n)}$, let $y+\{i,j\}$ denote the network whose set of edges is the union of the set of edges of $y$ and $\left\{ \{ i,j\} \right\}$ and let $y-\{i,j\}$ denote the network whose set of edges is the set difference between the set of edges of $y$ and $\left\{ \{ i,j\} \right\}$.  The \emph{change statistic} of $\mathbb{P}_\theta$ for $\{i,j\}$ is defined by
\begin{equation}\label{eqn:change_stat}
    \Delta_{i,j} g(y) := g\left(y + \{i,j\}\right) - g\left(y-\{i,j\}\right) \,. 
\end{equation}
The change statistic $\Delta_{i,j} g(y)$ determines the probability of occurrence of the edge $\{i,j\}$. More specifically, we have that 
\begin{equation} \label{eqn:edge_prob}
    \mathbb{P}_{\theta}\left(Y_{i,j}=1| Y-\{i,j\}=y-\{i,j\}\right) = \logit^{-1}\left(\theta^{\T} \Delta_{i,j}g(y)\right) \,,
\end{equation}
where $\logit^{-1}(x) :=1/(1+e^{-x})$ \cite{hunter2006inference}. For example, if $g_A$ is the homophily of a set of nodes $A$, then $\Delta_{i,j} g_A(y)$ is $1$ if $\{i,j\} \subseteq A$ and $0$ otherwise. Consequently, if a statistic vector $g$ contains $g_A$, then the log-odds of the conditional probability in (\ref{eqn:edge_prob}) contains the term $\theta_A \Delta_{i,j} g_A(y)$ and $\theta_A$ is positively correlated to the probability of occurrence of the edge $\{i,j\}$.

\begin{table}[]
    \centering
    \begin{tabular}{lll}
       {\bf Statistics}  & {\bf Name} & {\bf Description} \\ \hline \hline
       && \\
       $g_1$, $g_2$, $g_3$ &  Propensity of age group $i$,  & Number of times that a node in age group  \\
       & for $i=1,2,3$ & $i$ appears in an edge, where $i=1,2,3$\\
       && \\
       $g_4$, $g_5$ & Propensity of race group $i$, & Number of times that a node with race group  \\
       &  for $i \in \{ \textnormal{white, black}\}$ & $i$ appears in an edge, for $i \in \{\textnormal{white, black}\}$\\
       && \\
       $g_6$ & Propensity of PrEP users & Number of times that a PrEP node  \\
       & & appears in an edge \\
       && \\
       $g_7$ & Propensity of people  & Number of times that a person \\
       & who traded sex & who traded sex appears in an edge \\
       && \\
       $g_8$ & Propensity of people & Number of times that a person \\
       &  who had group sex & who had group sex appears in an edge \\
       && \\
       $g_9$, $g_{10}$, $g_{11}$, $g_{12}$ & Homophily of age group $i$,  & Number of edges with both nodes \\
       & for $i=1,2,3,4$ & in age group $i$, for $i=1,2,3,4$  \\
       && \\
       $g_{13}$, $g_{14}$, $g_{15}$ & Homophily of race group $i$, & Number of edges with both nodes in \\
       & for $i \in \{ \textnormal{white, black, other}\}$ & race group $i$, for $i \in \{\textnormal{white, black, other}\}$\\
       && \\
       $g_{16}$, $g_{17}$ & Homophily of no-PrEP and  & Number of edges with both nodes in group $i$, \\
       & PrEP group &  for $i \in \{ \textnormal{no-PrEP group, PrEP group} \}$ \\
       && \\
       $g_{18}$, $g_{19}$, $g_{20}$ & Monogamy of race group $i$,  & Number of nodes with degree one in\\
       & for $i \in \{\textnormal{white, black, other}\}$ &  race group $i$, for $i \in \{\textnormal{white, black, other}\}$  \\
       && \\
       \hline \hline 
       
    \end{tabular}
    \caption{Statistics used in the fitted linear ERGM. }
    \label{tab:stats_fitted}
\end{table}

Let $A$ be a group of nodes. The {\it propensity of $A$} is the number of times that a node in $A$ appears in an edge, the {\it homophily} of $A$ is the number of edges with both nodes in $A$, and the {\it monogamy} of $A$ is the number of nodes in $A$ with degree 1. Previous work has identified propensity, homophily, and monagamy of racial groups as important features of MSM sexual networks \cite{janulis2018sexual}. Consequently, we include the statistics associated with these features in the linear ERGM that we fit.  Age is associated with infection disparities \cite{CDC_syphilis}, so we also include age-related statistics in the ERGM.  Specifically, we consider four age groups ([18,25), [25,35), [35,45), and 45+), and include propensity and homophily of these age groups in the fitted linear ERGM.  We also examine the significance of PrEP usage in the probability of partnership formation by including the propensity and homophily of PrEP users in the model. The largest degree among the egos in the sample $e_{\mathbf{S}}$ was 62 during 2019 and 2020. To fit the observed large degree for some egos, we include the propensity of MSM who traded sex and the propensity of MSM who had group sex. Altogether this yields 20 statistics $g_i$ for the ERGM, listed in Table \ref{tab:stats_fitted}.

Let $\tilde{\theta} =\left(\theta_1,\ldots,\theta_{20}\right)^{\T}$ denote the coefficient vector that is estimated from the NEST data.  
We estimate $\tilde{\theta}$ from the egocentric sample $\es$ (and sample weights\footnote{The sample weights $\tilde{w}_i$ compensate biases due to sampling (see Appendix \ref{sec:egocentric_samples}). We compute these sample weights taking into account the age and race distributions of the data and the census of Franklin county (see Appendix \ref{sec:ERGM_NEST}).} $\tilde{w}_i$, for $i \in \mathbf{S}$) using the method of pseudo maximum likelihood estimation (PMLE). See Appendix \ref{sec:fitting_ERGMs} for more details on PMLE, and see Table \ref{tab:fitted_ergm} in Appendix \ref{sec:ERGM_NEST} for details on the estimate $\tilde{\theta}$. The coefficients that are significantly different from zero [{and thus have} an effect on the edge occurrence of the ERGM by equation (\ref{eqn:edge_prob})] are the propensity and homophily of age groups, the homophily and monogamy of racial groups, and the propensity of people having group sex. The coefficients of propensity and homophily of PrEP users, propensity of racial groups, and propensity of people who traded sex are not significantly different from zero.  We give the coefficient estimates and associated p-values in Table \ref{tab:fitted_ergm} of Appendix \ref{sec:ERGM_NEST}.

\subsubsection{ERGM fit to NEST data} \label{sesc:descriptive_ergm_data}

Let $\tilde{y}_{(1)},\ldots, \tilde{y}_{(100)}$ be  sampled networks from the fitted linear ERGM $\mathbb{P}_{\tilde{\theta}}$ to the NEST egocentric sample $e_{\mathbf{S}}$. Each network $\tilde{y}_{(k)}$ has 2978 nodes. The age, race, PrEP usage, sex-trade, and group-sex attributes are drawn from the same distribution for all networks (see Table \ref{tab:descriptive_ergm_data}). All the statistics that we give throughout the rest of the paper, such as means and standard deviations (\textit{sd}), are computed over the largest connected components of the sampled networks $\tilde{y}_{(1)},\ldots, \tilde{y}_{(100)}$. The largest connected component $y_{\textnormal{ERGM}(k)}$ of $\tilde{y}_{(k)}$ has a mean size of 2954 $(sd = 4.92)$.  We use the 100 networks $y_{\textnormal{ERGM}(k)}$ for our analysis of community structure informed by network connectivity  and PrEP usage {(Section \ref{sec:community})} and for simulating syphilis dynamics {(Section \ref{sec:dynamics})}.

In the fourth column of Table \ref{tab:descriptive_ergm_data}, we show the attribute distributions of $y_{\textnormal{ERGM}(k)}$ obtained after using sample weights to compensate for biases due to sampling in the NEST study. {We observe that the mean percent of nodes of $y_{\textnormal{ERGM}(k)}$ that are PrEP nodes is 28.1\% ($sd = 0.05\%$).} We use the age and race distributions of the {2020} census data {for} Franklin county as reference for our sample weights. Note that we do not have demographic information for the entire MSM population in Franklin county. Further details are given in Appendices \ref{sec:egocentric_samples} and \ref{sec:ERGM_NEST}.

\begin{table}[H]
\centering
\begin{tabular}{llll}
{\bf Attribute} & {\bf Attribute value} & $\boldsymbol{e_{\mathbf{S}}}$ & $\boldsymbol{y}_{\textnormal{ERGM}(k)}$ \\
 \hline \hline

{\bf Age} & Group 1: age < 25 & 23.3\%  & $ 14.7 \%$ $ (sd = 0.05\%)$ \\
& Group 2: 24 < age < 35 & 43.2\%  & $ 38.0 \% $ $(sd = 0.05\%)$ \\
& Group 3: 34 < age < 45 & 16.1\%  & $ 18.9 \% $ $ (sd = 0.04\%)$ \\
& Group 4: 44 < age & 17.4\%  & $ 28.4 \% $ $(sd = 0.1\%)$ \\
\hline
{\bf Race} & white & 69.5\%  & $ 62.0\% $ $ (sd = 0.08\%)$ \\ 
& black & 24.2\%  & $ 23.2 \% $ $ (sd = 0.1\%)$ \\ 
& other & 6.3\%  & $ 14.8\% $ $ (sd = 0.1\%)$ \\ 
\hline
{\bf PrEP usage} & yes & 32.6\% & $28.1 \% $ $ (sd = 0.05\%)$ \\
& no & 67.4\% & $ 71.9 \% (sd = 0.05\%)$ \\
\hline 
{\bf Sex trade} & yes & 7.6\% & $7.7 \% $ $ (sd = 0.04\%)$ \\
& no & 92.4\% & $92.3 \% $ $ (sd = 0.04\%)$ \\
\hline
{\bf Group sex} & yes & 40.3\% & $38.0 \% $ $ (sd = 0.06\%)$ \\
& no & 59.7\% & $62.0 \% $ $ (sd = 0.06\%)$ \\
\hline
\hline
\end{tabular}
\caption[]{Attribute distributions for the egocentric sample network $e_{\mathbf{S}}$ and the networks $y_{\textnormal{ERGM}(k)}$.}
\label{tab:descriptive_ergm_data}
\end{table}

\begin{figure}[H]
\centering
\subfloat[]{\includegraphics[scale = 0.4]{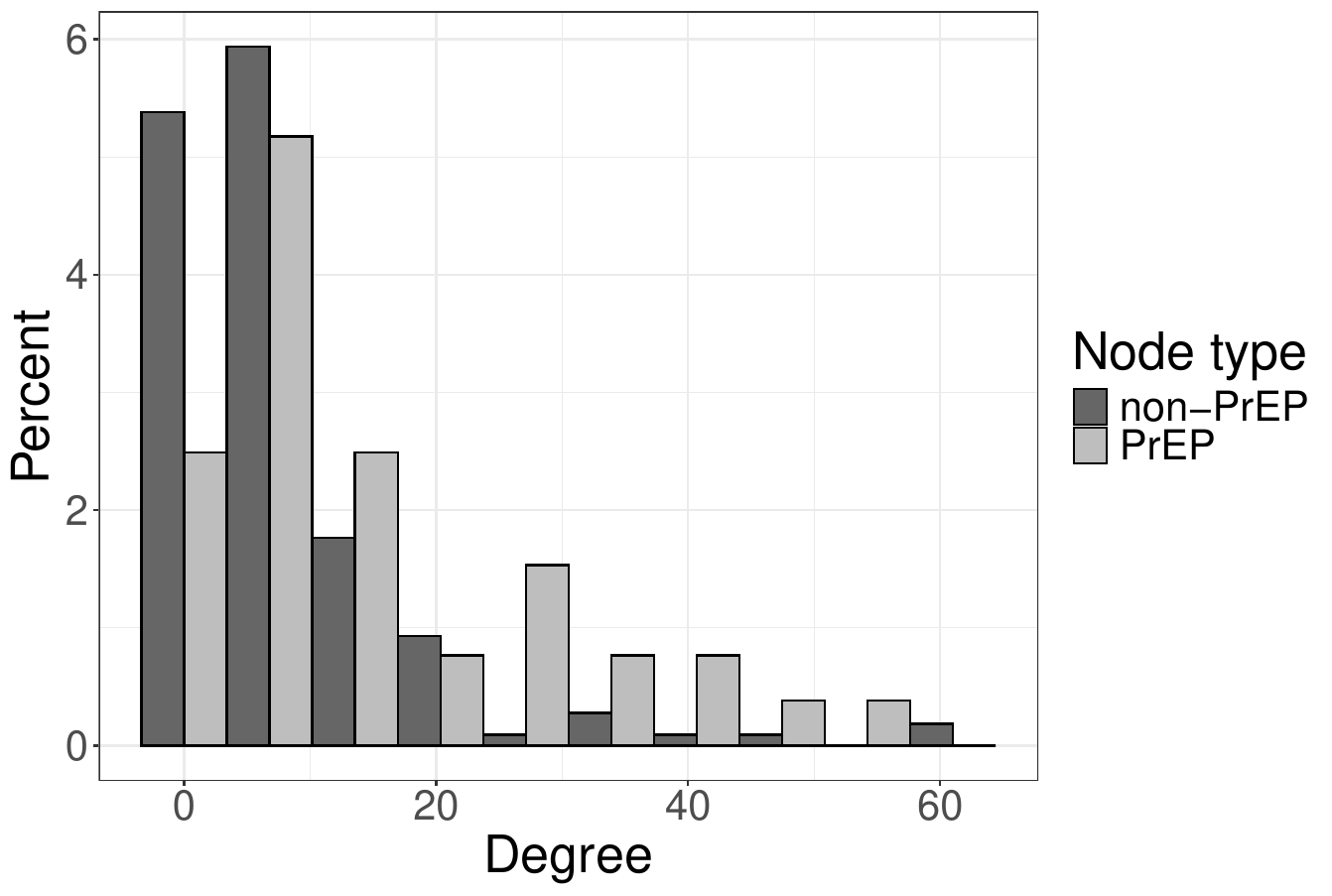}}
\subfloat[]{\includegraphics[scale = 0.4]{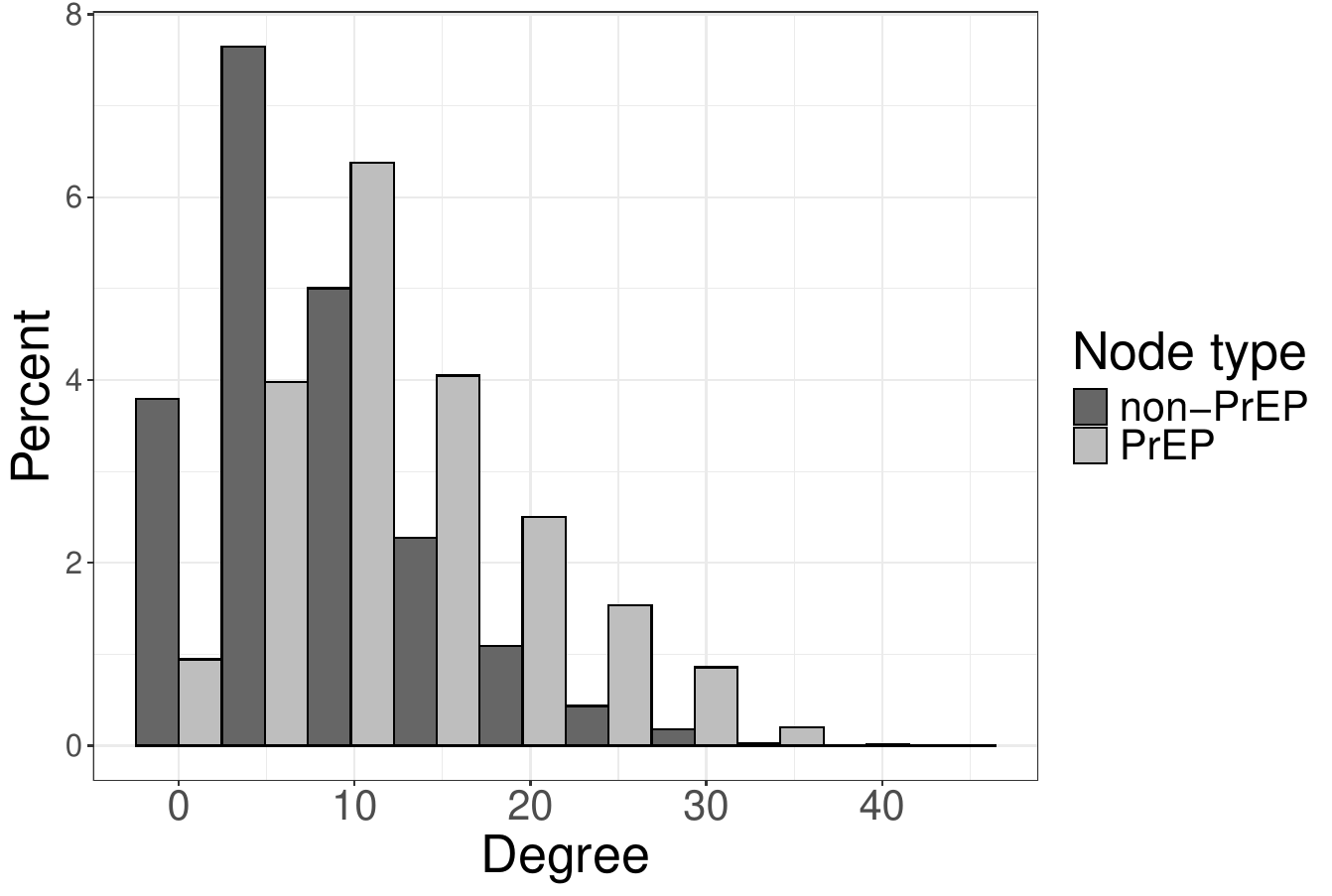}} 
\caption[]{Degree distributions conditioned to PrEP and non-PrEP nodes for the egocentric sample $e_{\mathbf{S}}$ and averaged over the networks $y_{\textnormal{ERGM}(k)}$, $k=1,\dots,100$. (a) Degree distributions conditioned to PrEP and non-PrEP egos for $e_{\mathbf{S}}$. (b) Degree distributions conditioned to PrEP and non-PrEP nodes for $y_{\textnormal{ERGM}(k)}$.}
\label{fig:degree_by_tr}
\end{figure}

We find that the mean degree of $y_{\textnormal{ERGM}(k)}$ is 9.2 $(sd = 6.7)$. We also observe that on average the PrEP nodes of $y_{\textnormal{ERGM}(k)}$ have larger degree than the non-PrEP nodes of $y_{\textnormal{ERGM}(k)}$. Specifically, the conditional mean degree for PrEP nodes of $y_{\textnormal{ERGM}(k)}$ is 13.0 $(sd = 7.28)$ and the conditional mean degree for non-PrEP nodes of $y_{\textnormal{ERGM}}$ is 7.98 $(sd = 6.05)$. In Figure \ref{fig:degree_by_tr}, we show the degree distributions by PrEP status for $e_S$ (Figure \ref{fig:degree_by_tr}(a)) and $y_{\textnormal{ERGM}(k)}$ (Figure \ref{fig:degree_by_tr}(b)).
In Figure \ref{fig:degree_by_tr}(b), we observe that the degree distribution of the networks $y_{\textnormal{ERGM}(k)}$ conditioned to PrEP nodes is heavier to the right than the corresponding degree distribution conditioned to non-PrEP nodes.  This is consistent with the conditioned degree distributions of $\es$ in Figure \ref{fig:degree_by_tr}(a). The mean monogamy proportion in $y_{\textnormal{ERGM}(k)}$ is 11.1\% $(sd = 0.525\%)$, which matches the monogamy proportion of $e_{\mathbf{S}}$; this is due to the inclusion of the statistics of monogamy of race groups in $\mathbb{P}_{\tilde{\theta}}$.

\section{Syphilis transmission model}\label{sec:syphilis_model_3}
\begin{figure}[H]
    \centering
    \subfloat[]{\includegraphics[scale=0.25]{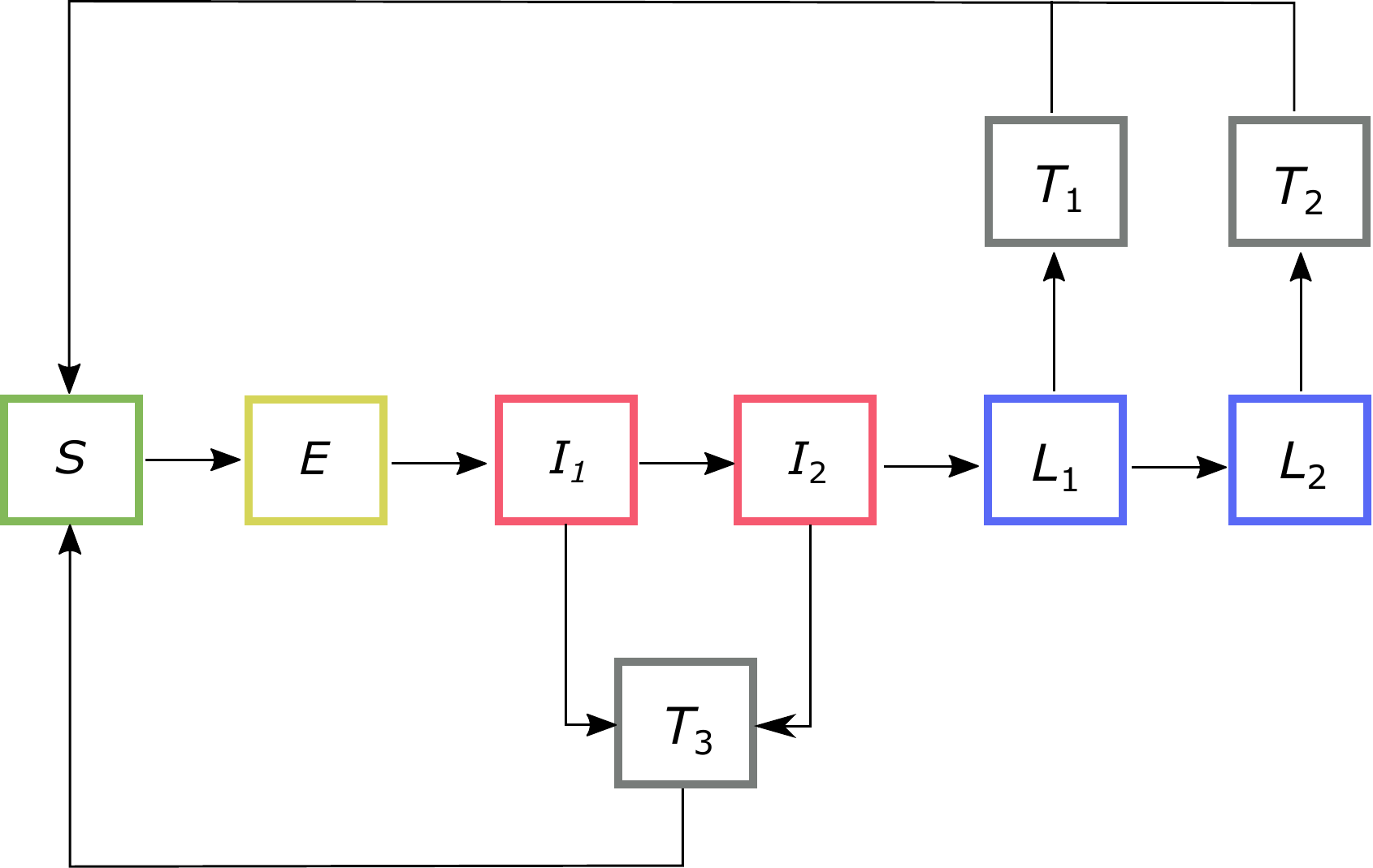}} \quad
    \subfloat[]{\includegraphics[scale = 0.25]{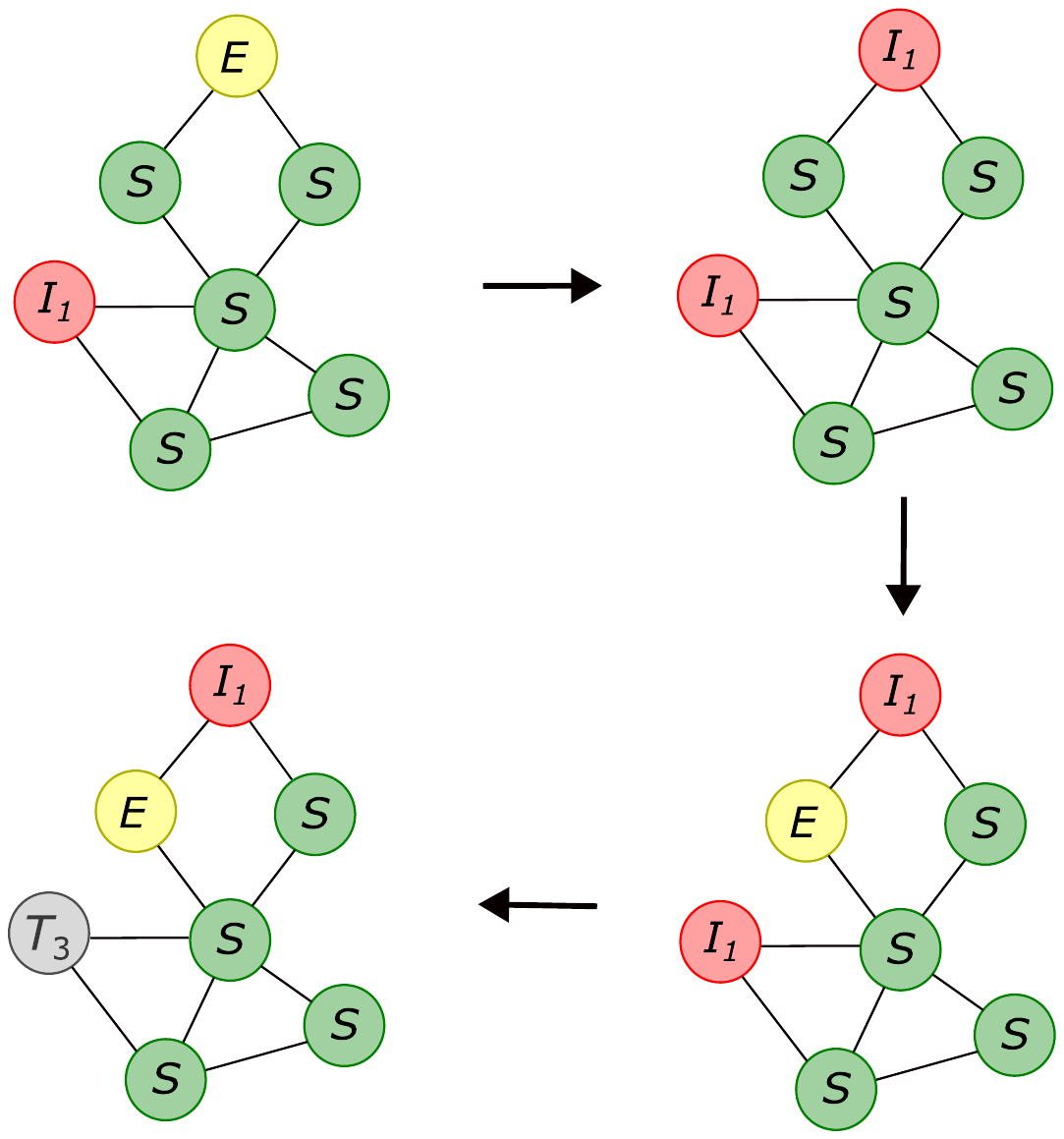}}  
    \caption{
    Network schematics. (a) State transitions for a single node of $y_{\textnormal{ERGM}(k)}$ in our stochastic syphilis model. (b) Example of possible changes in node status over time under our syphilis model on a mock network. }
    \label{fig:transition_syphilis}
\end{figure}

The syphilis model that we use is a stochastic model simulating syphilis dynamics on MSM sexual networks. 
The states and transitions of our stochastic network model follow those of the differential equation model of Tuite and Fisman \cite{tuite2016go}. Nodes in the network are categorized according to the following possible states with respect to syphilis: susceptible ($S$), exposed ($E$), primary syphilis ($I_1$), secondary syphilis  ($I_2$), early latent syphilis ($L_1$), and late latent syphilis ($L_2$).  Treated people recover from infection and are assumed to be temporarily immune to re-infection, with $T_1, T_2$ corresponding to temporarily immune people following treatment during early latent and late latent syphilis, respectively, and $T_3$ corresponding to temporarily immune people following treatment during primary or secondary syphilis. As in \cite{tuite2016go}, we assume that only people in primary or secondary syphilis are infectious, and we neglect superinfection by multiple syphilis strains. A flow diagram showing possible node state transitions is shown in Figure \ref{fig:transition_syphilis}(a), and a schematic showing an example of disease dynamics under the model on a toy network is shown in Figure \ref{fig:transition_syphilis}(b).

We use a continuous-time Markov process to model syphilis dynamics on the largest connected components $y_{\textnormal{ERGM}(k)}$ of the sampled networks $\tilde{y}_{(k)}$ from the fitted linear ERGM. We simulate the model using the Gillespie algorithm \cite{gillespie1976general,kiss2017, pastor2015epidemic}. The state transitions and rate parameters for our continuous-time Markov process are given in Table \ref{tab:syphilis_rates_nest}.  New infections occur via transmission along an edge connecting a susceptible ($S$) and infectious ($I_1$ or $I_2$) node with rate parameter $\beta$.  Following infection, people enter an exposed, non-infectious state ($E$). Exposed people become infectious and enter primary syphilis ($I_1$) with rate parameter $\gamma_0$.  People transition from primary syphilis to secondary syphilis ($I_2$) at rate $\gamma_1$, from secondary syphilis to early latent syphilis ($L_1$) at rate $\gamma_2$, and from early latent to late latent syphilis ($L_2$) at rate $\gamma_3$.    

People in our model remain infected until they are treated, upon which they enter one of three treated states ($T_1$ for treatment during early latent syphilis, $T_2$ for treatment during late latent syphilis, and $T_3$ for treatment during primary or secondary syphilis).  The expected immune duration for a person in $T_i$ is given by $1/\lambda_i$.  Rate parameters are based on \cite{tuite2016go} and listed in Table \ref{tab:syphilis_rates_nest}.  
Let $\alpha_1, \alpha_2, \alpha_3, \alpha_4$ be the per capita treatment rates for people in $I_1, I_2, L_1, L_2$, respectively.  We decompose $\alpha_i$ into several parts, including a background rate due to antibiotic treatment for non-syphilis infections, and a treatment rate due to syphilis screening programs.  This latter treatment rate is higher for people using PrEP ($\alpha + \alpha_{\textnormal{PrEP}}$, with $\alpha_{\textnormal{PrEP}}>0$) versus not using PrEP ($\alpha + \alpha_{\textnormal{PrEP}}$, with $\alpha_{\textnormal{PrEP}} = 0$).  We refer to $\alpha$ as the {\it baseline treatment rate.} 
We assume that PrEP users are tested for syphilis every three months due to associated visits to a sexual health provider. We also assume that  the probability of getting infected is uniform over the three-month period between tests.  Under these assumptions the expected time to treatment due to PrEP is 1.5 months, giving $\alpha_{\textnormal{PrEP}} = 1/1.5$ for PrEP nodes.
We additionally include a treatment rate due to symptom-driven health-care seeking.  This latter rate depends upon the syphilis state of the infected person.

\begin{table}[H]
    \centering
    \begin{tabular}{lll}
    \textbf{Transition}     &  \textbf{Rate} & \textbf{Parameters (per month)} \\
    \hline \hline 
    {\it Infection} & & \\
    $ |{\rm S}| \rightarrow |{\rm S}| -1, \quad |{\rm E}| \rightarrow |{\rm E}|+1$     & $\beta |{\rm SI}| $ & $\beta$ varies \\
    & & \\
    {\it Syphilis progression} & & \\  
    $ |{\rm E}| \rightarrow |{\rm E}| - 1, \quad  |{\rm I_1}| \rightarrow |{\rm I_1}| +1$ & $\gamma_0 |{\rm E}| $ & $\gamma_0 = 1/0.9$ \\
    $ {\rm |I_1| \rightarrow |I_1| -1, \quad |I_2| \rightarrow |I_2| +1}$ & $\gamma_1 |{\rm I_1}|$ & $\gamma_1 = (1/1.5)\times 0.85 $\\
    ${\rm |I_2| \rightarrow |I_2| -1, \quad |L_1| \rightarrow |L_1| +1}$ & $\gamma_2 |{\rm I_2}|$ & $\gamma_2 = (1/3.6)\times 0.75$ \\
    ${ \rm |L_1| \rightarrow |L_1| -1, \quad  |L_2| \rightarrow |L_2| +1 }$ & $\gamma_3 |{\rm L_1}|$ & $\gamma_3 = (1/6.9) \times 0.75 $ \\
    & & \\
    {\it Treatment} & & \\
    ${ \rm |I_1| \rightarrow |I_1| -1, \quad |T_3| \rightarrow |T_3| + 1}$ & $\alpha_1 |{\rm I_1}|$ & $\alpha_1 = [(1/1.5) \times 0.15] + 0.001/12 + \alpha + \alpha_{\textnormal{PrEP}} $, $\alpha$ varies\\
    ${\rm |I_2| \rightarrow |I_2| - 1, \quad |T_3| \rightarrow |T_3| + 1}$ & $\alpha_2 |{\rm I_2}|$ & $\alpha_2 = [(1/3.6) \times 0.25] + 0.001/12 + \alpha + \alpha_{\textnormal{PrEP}}$, $\alpha$ varies\\
    ${\rm |L_1| \rightarrow |L_1| - 1, \quad  |T_1| \rightarrow |T_1| +1 }$ & $\alpha_3 |{\rm L_1}|$ & $\alpha_3 = [(1/6.9) \times 0.25] + 0.001/12 + \alpha + \alpha_{\textnormal{PrEP}}$, $\alpha$ varies\\
    ${ \rm |L_2| \rightarrow |L_2| -1, \quad  |T_2| \rightarrow |T_2| +1}$ & $\alpha_4 |{\rm L_2}|$ & $\alpha_4 = 0.001/12 + \alpha + \alpha_{\textnormal{PrEP}} $, $\alpha$ varies \\
    & & \\
{\it Loss of immunity} & & \\	
    ${\rm |T_3| \rightarrow |T_3| -1, \quad |S| \rightarrow |S| +1}$ & $\lambda_3 |{\rm T_3}|$ & $\lambda_3 = 1/0.25$\\
    ${\rm |T_1| \rightarrow |T_1| -1, \quad |S| \rightarrow |S| +1}$ & $\lambda_1 |{\rm T_1}|$ & $\lambda_1 = 1/0.25$\\
    ${\rm |T_2| \rightarrow |T_2| -1, \quad |S| \rightarrow |S| +1}$ & $\lambda_2 |{\rm T_2}|$ & $\lambda_2 = 1/60$\\
    
    \hline
    \hline
    \end{tabular}
    \caption[Transition rates of the syphilis model used on the NEST network]{Transition rates in the model for the spread of syphilis on a network. The baseline treatment rate $\alpha$ and the transmission rate $\beta$ are varied, and the rate $\alpha_{\textnormal{PrEP}}$ is $1/1.5$ for infectious people who use PrEP and zero otherwise. We take the parameter values from 
     \cite{tuite2016go} for the other rates in this table. The parameter values give monthly rates. }
    \label{tab:syphilis_rates_nest}
\end{table}

\section{Community structure, treatment rates, and PrEP}\label{sec:community_fitted_ERG}
\label{sec:community}
\begin{figure}[H]
    \centering
    {\includegraphics[scale = 0.4]{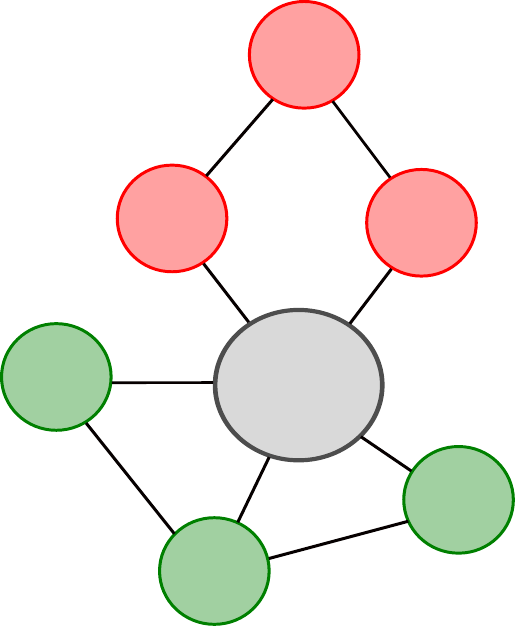}} 
    \caption{
    Schematic showing how differential disease treatment rates can influence effective community structure.  Let the large grey node have a larger disease treatment rate than the other nodes.  The grey node then serves as a barrier for disease spread from the red nodes to the green nodes, effectively separating the green and red nodes into different communities.  } 
    \label{fig:large_vs_small_abs}
\end{figure}

We now consider the community structure of the sampled networks from the fitted ERGM.  In particular, we perform community detection on the largest connected components $y_{\textnormal{ERGM}(k)}$ of the sampled networks for $k = 1, \dots, 100$. We use InfoMap for absorbing random walks  \cite{vargas2022infomap} to perform community detection in a manner that takes both sexual contacts and disease treatment rates into account. InfoMap is a widely-used community detection algorithm \cite{rosvall2009map,rosvall2008maps} based upon random walks. Standard InfoMap uses regular random walks, where there are no absorbing states and the random walk converges to a unique stationary distribution with positive probabilities for every node.  A random walker will tend to spend long periods of time within tightly-knit clusters of nodes, corresponding to the communities detected by InfoMap.  In the context of disease spread, treatment rates of the disease may influence the effective community structure of a network.  For example, a node with low treatment rate can effectively increase connectivity between different portions of the network, while a node with high treatment rate inhibits the spread of disease and can effectively separate subsets of nodes. See Figure \ref{fig:large_vs_small_abs} for a schematic. This can lead to effective community structure that differs from community structure detected using methods based upon network structure alone.

{Here we use PrEP status as a proxy for syphilis screening rates under the rationale that people on PrEP are likely in regular contact with a sexual health provider \cite{tang2020quarterly}.} {We assume that syphilis treatment rates for people on PrEP are higher than for people not on PrEP.}  Given a sampled network $y_{\textnormal{ERGM}(k)}$ with nodes $1, \dots, N_k$ and adjacency matrix $A_{\textnormal{ERGM}(k)}$, let $\delta_i$ be the \emph{node-absorption rate} of node $i$.  We interpret the flow of the disease as a random walk that transitions according to the contacts in the sexual network (specified by  $A_{\textnormal{ERGM}(k)}$) and is absorbed according to treatment, where the treatment rate is associated with the $\delta_i$.  We use InfoMap for absorbing random walks to obtain partitions $M$ of the set of nodes $\{1, \ldots, {N_k}\}$ that reflect the effect of the node-absorption rates $\delta_1, \ldots, \delta_{{N_k}}$ on the community structure of $y_{\textnormal{ERGM}(k)}$.  Further details are given in Appendix \ref{appendix:infomap} and \cite{vargas2022infomap}. We refer to a partition that captures the effect of absorption on the community structure of $y_{\textnormal{ERGM}(k)}$ as \emph{effective community structure} of $y_{\textnormal{ERGM}(k)}$. 

In our subsequent analysis we use two possible values for the node-absorption rates: $\delta_{\textnormal{PrEP}}$ for PrEP nodes and $\delta_{\textnormal{no-PrEP}}$ for non-PrEP nodes.  We assume that $$\delta_{\textnormal{no-PrEP}} < \delta_{\textnormal{PrEP}}\,,$$
reflecting more frequent syphilis treatment for PrEP versus non-PrEP nodes.  We parameterize $\delta_{\textnormal{no-PrEP}}$  and $\delta_{\textnormal{PrEP}}$ to be consistent with the syphilis model in Section \ref{sec:syphilis_model_3}, where the mean time that an untreated person remains infectious is $5.1$ months ($1.5$ months for the primary syphilis state $I_1$, and $3.6$ months for the secondary syphilis state $I_2$), and the mean time until  a PrEP user receives treatment is $1.5$ months.  Scaling the node-absorption rates relative to the treatment rate for PrEP nodes gives 
\begin{align}
    \delta_{\textnormal{PrEP}} := 1, \quad \text{and} \quad \delta_{\textnormal{no-PrEP}} := \frac{1/5.1}{(1/5.1) + (1/1.5)} \,.
\end{align}

We consider two absorption rate scenarios. In the first scenario (that we refer as \emph{no-PrEP scenario}), we treat all the nodes in $y_{\textnormal{ERGM}(k)}$ as non-PrEP nodes and define the entries of the node-absorption-rate vector $\vec{\delta}_{\textnormal{no-PrEP}}$ by $\left(\vec{\delta}_{\textnormal{no-PrEP}}\right)_i := \delta_{\textnormal{no-PrEP}}$ for $i \in \{1,\ldots,{N_k}\}$. In the second scenario (that we refer as \emph{PrEP scenario}), we define the entries of the node-absorption-rate vector $\vec{\delta}_{\textnormal{PrEP}}$ by $\left(\vec{\delta}_{\textnormal{PrEP}}\right)_i := \delta_{\textnormal{PrEP}}$ if $i$ is a PrEP node in $y_{\textnormal{ERGM}(k)}$ and $\left(\vec{\delta}_{\textnormal{PrEP}}\right)_i := \delta_{\textnormal{no-PrEP}}$ otherwise. Table \ref{tab:absorption_scenarios} summarizes the proportions of nodes that are treated as PrEP nodes in each of the two absorption scenarios. 

\begin{table}[H]
    \centering
    \begin{tabular}{lll}
    {\bf Absorption scenario}   & {\bf Node-absorption-rate}  & {\bf Percent of nodes that } \\
     & {\bf vector } & {\bf are treated as PrEP nodes}\\ 
     \hline \hline
     no-PrEP scenario & $\vec{\delta}_{\textnormal{no-PrEP}}$ & 0\% \\
     PrEP scenario & $\vec{\delta}_{\textnormal{PrEP}}$ & 28.1\% $(sd = 0.05\%)$ \\
     \hline \hline
    \end{tabular}
    \caption{Absorption-rate scenarios. A node is treated as a PrEP node if it has node-absorption rate $\delta_{\textnormal{PrEP}}$ (or $\alpha_{\textnormal{PrEP}} = 1/1.5$).}
    \label{tab:absorption_scenarios}
\end{table}

\begin{figure}[H]
    \centering
    \subfloat[]{\includegraphics[scale = 0.45]{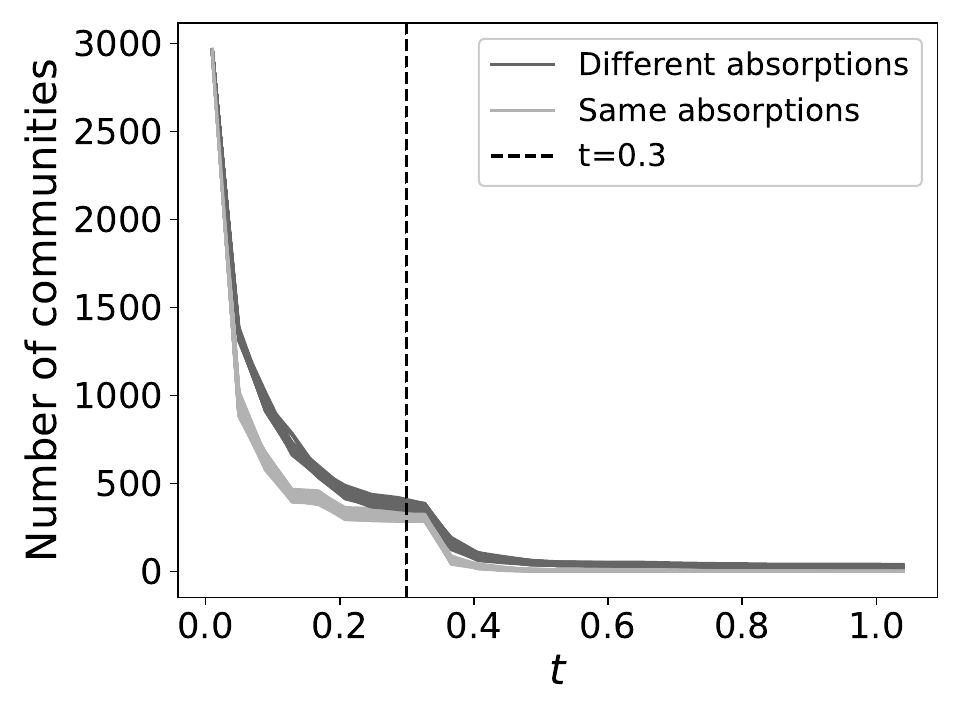}}
    \subfloat[]{\includegraphics[scale = 0.45]{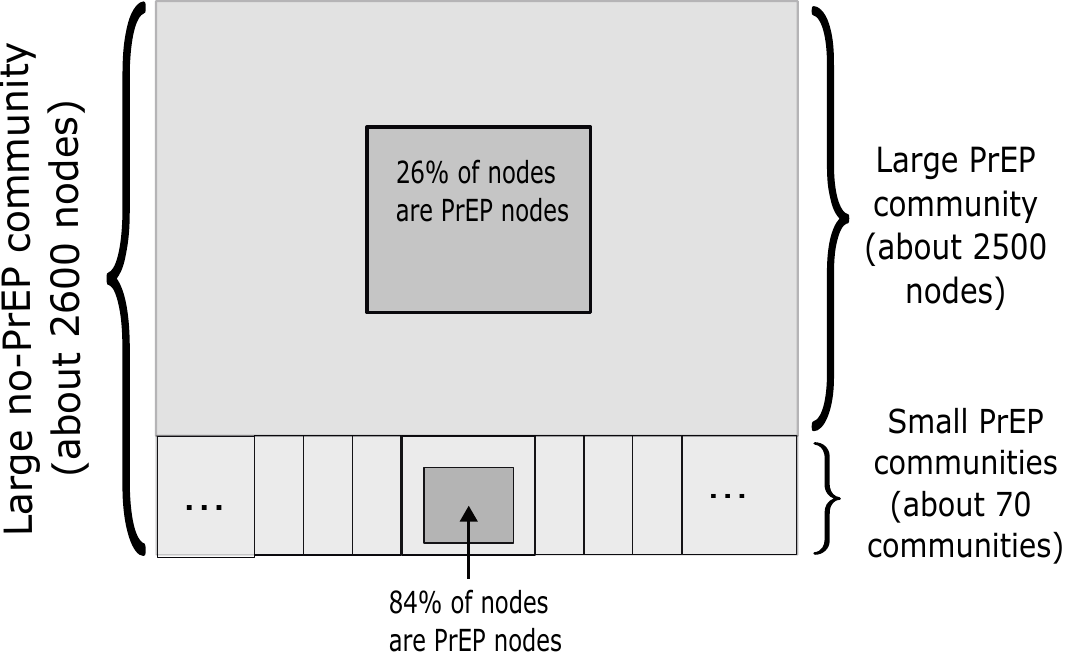}}
    \caption[]{Effective community structure of $y_{\textnormal{ERGM}(k)}$ for different scenarios of absorption rates. (a) Number of communities of $y_{\textnormal{ERGM}(k)}$. Number of communities obtained from InfoMap for absorbing random walks in Algorithm \ref{alg:adapt_info} of Appendix \ref{appendix:infomap} with inputs $A_{\textnormal{ERGM}(k)}$, for $k=1,\ldots,100$, $\vec{\delta} \in \left\{ \vec{\delta}_{\textnormal{no-PrEP}},  \vec{\delta}_{\textnormal{PrEP}^-} \right\}$, and Markov time $t$. (b) Splitting effect of PrEP usage. Under the PrEP scenario, the large no-PrEP community of $y_{\textnormal{ERGM}(k)}$ is split into the large PrEP community of $y_{\textnormal{ERGM}(k)}$ and small PrEP communities of $y_{\textnormal{ERGM}(k)}$. We observe that on average 26\% of the nodes in the large PrEP community of $y_{\textnormal{ERGM}(k)}$ are PrEP nodes, while on average 84\% of the nodes in the small PrEP communities of $y_{\textnormal{ERGM}(k)}$ are PrEP nodes.}
    \label{fig:markov_times}
\end{figure}

In Figure \ref{fig:markov_times}, we show the number of communities detected for $y_{\textnormal{ERGM}(k)}$ in the PrEP and no-PrEP scenarios using InfoMap for absorbing random walks.  The horizontal axis is a resolution parameter for InfoMap called Markov time $t$ \cite{schaub2012encoding}. This parameter tunes the sizes of the yielded communities by InfoMap for absorbing random walk, where small values of $t$ produce small communities.  There is a range of Markov time values near $t=0.3$ for which the number of communities changes little as Markov time is varied, reflecting robustness with respect to the resolution parameter.  Examining the detected communities for $t=0.3$, we find that the largest community of $y_{\textnormal{ERGM}(k)}$ for the no-PrEP scenario (that we call \emph{the large no-PrEP community of $y_{\textnormal{ERGM}(k)}$}) is split into the set of nodes in the largest community of $y_{\textnormal{ERGM}(k)}$ for the PrEP scenario (that we call \emph{the large PrEP community of $y_{\textnormal{ERGM}(k)}$}) and the sets of nodes in small communities of $y_{\textnormal{ERGM}(k)}$ for the PrEP scenario (that we call \emph{small PrEP communities of  $y_{\textnormal{ERGM}(k)}$}) [see Figure \ref{fig:markov_times}(b)]. Specifically, the large no-PrEP community of $y_{\textnormal{ERGM}(k)}$ (that has a mean size of 2595 nodes)  
contains the large PrEP community (that has a mean size of 2513 nodes) 
and 67 small PrEP communities (on average) with sizes between one and three.  The communities used in the analysis of the subsequent sections are only the large and small PrEP communities of $y_{\textnormal{ERGM}(k)}$.

Most of the nodes in small PrEP communities of  $y_{\textnormal{ERGM}(k)}$ are PrEP nodes. Specifically, we observe that the proportion of nodes in the large PrEP community of $y_{\textnormal{ERGM}(k)}$ that are PrEP nodes is 25.7\% $(sd = 0.294\%)$, and the proportion of nodes in the small PrEP communities of $y_{\textnormal{ERGM}(k)}$ that are PrEP nodes is 83.7\% $(sd = 3.85\%)$. Consequently, a  significant proportion of the PrEP nodes in the large no-PrEP community of $y_{\textnormal{ERGM}(k)}$ are in small PrEP communities. Specifically, we observe that 9.6\% $(sd = 1.14\%)$ of the PrEP nodes in the large no-PrEP community of $y_{\textnormal{ERGM}(k)}$ belong to small PrEP communities.

\begin{figure}[H]
    \centering
    \subfloat[]{\includegraphics[scale = 0.4]{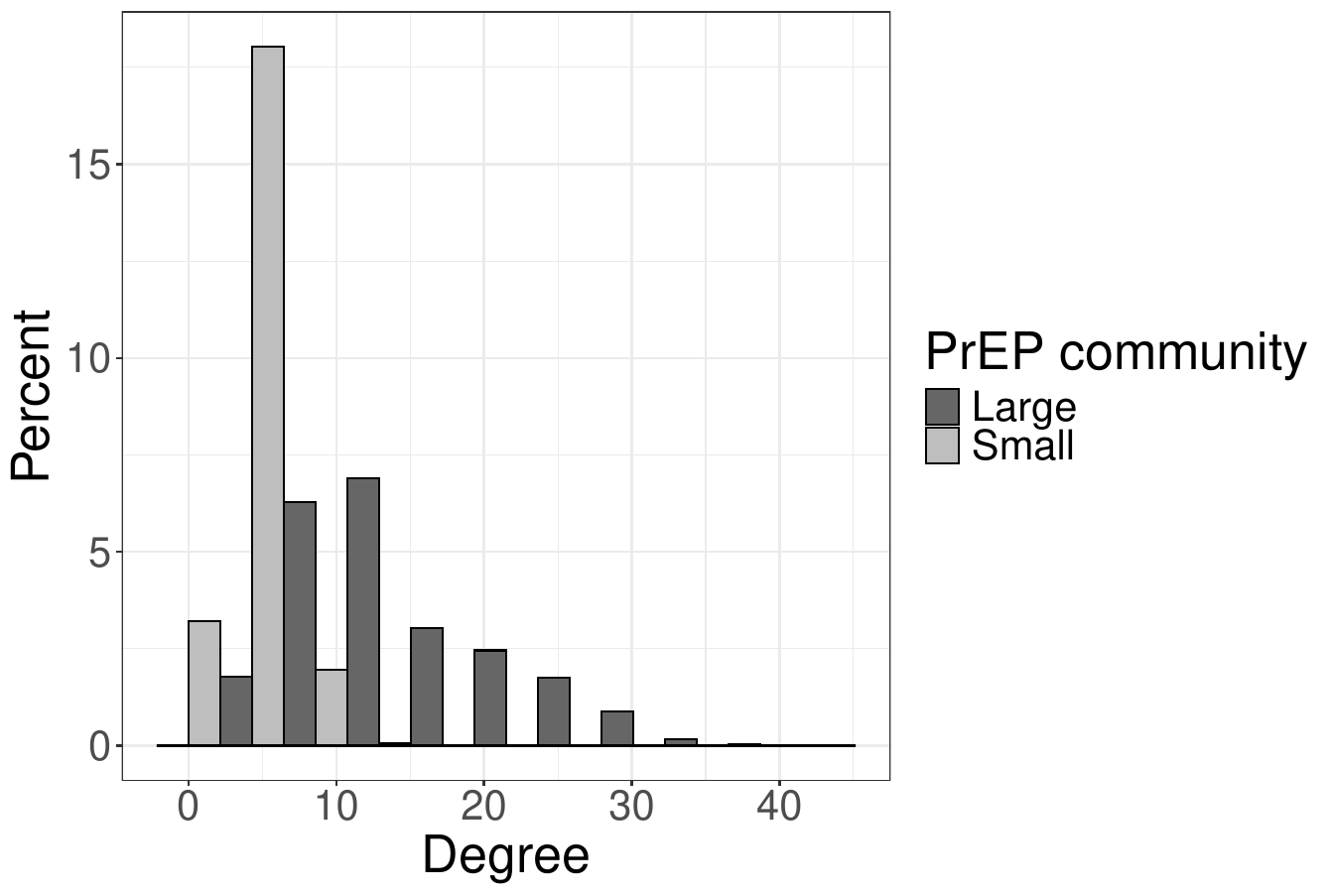}}
     \centering
    \subfloat[]{\includegraphics[scale = 0.4]{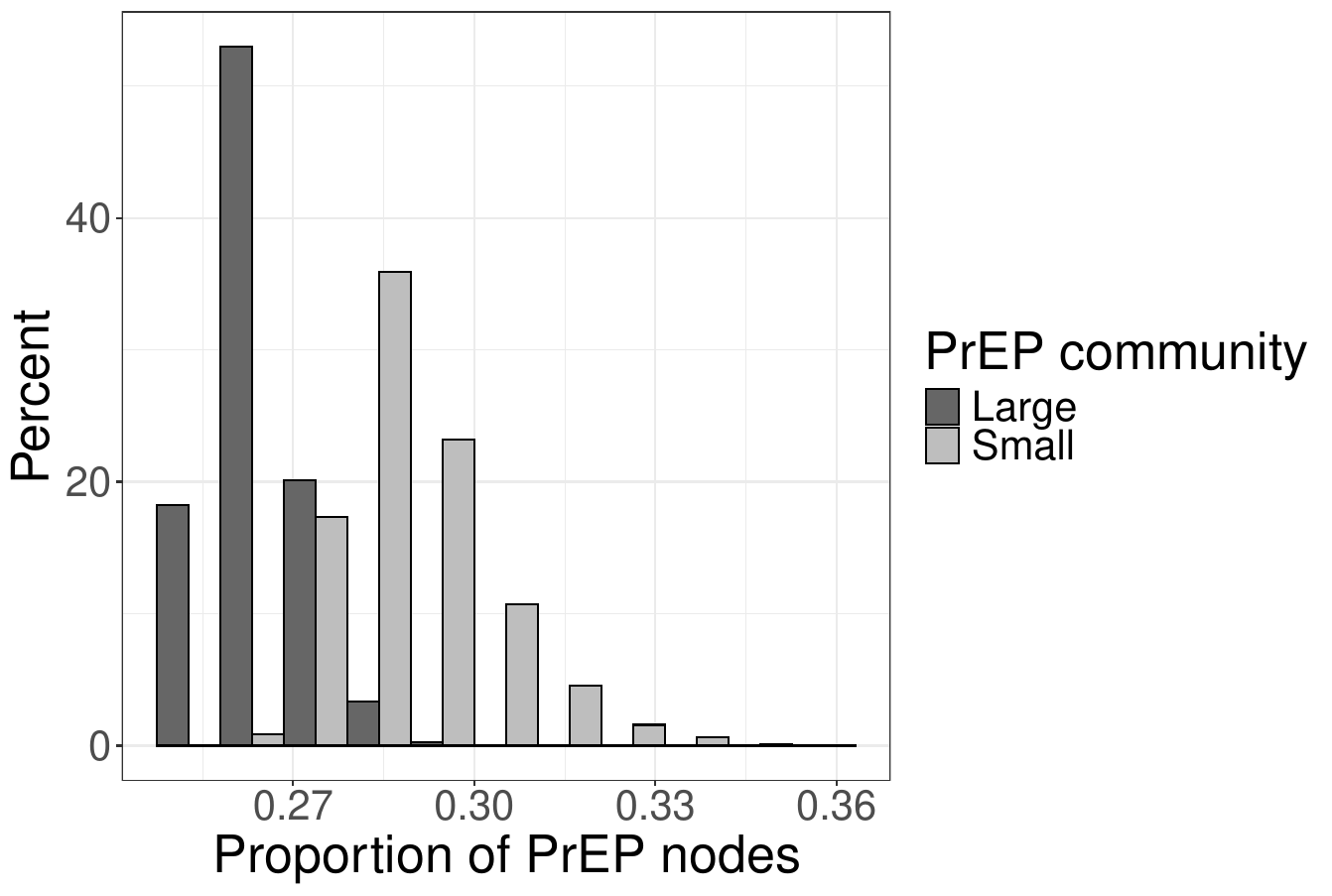}}
    \caption[]{Degree and proportion of PrEP neighbors of the PrEP nodes in the large and small PrEP communities of $y_{\textnormal{ERGM}(k)}$. (a) Degree distributions of the PrEP nodes in the large PrEP community of $y_{\textnormal{ERGM}(k)}$ (dark grey) and degree distribution of the PrEP nodes in small PrEP communities of $y_{\textnormal{ERGM}(k)}$ (light grey). (b) Distribution of the proportion of PrEP nodes in the neighborhoods of radius four of PrEP nodes in the large PrEP community of $y_{\textnormal{ERGM}(k)}$ (dark grey) and that distribution of the PrEP nodes in small PrEP communities of $y_{\textnormal{ERGM}(k)}$ (light grey). }
    \label{fig:small_comm_vs_large_comm}
\end{figure}

The large PrEP community of $y_{\textnormal{ERGM}(k)}$ contains PrEP nodes that have a relatively large degree and are connected to a relatively small number of PrEP neighbors. Specifically, we find that the PrEP nodes in the large PrEP community of $y_{\textnormal{ERGM}(k)}$ have a mean degree of 14.4 (with $sd = 6.67$), whereas the PrEP nodes in small PrEP communities of $y_{\textnormal{ERGM}(k)}$ have a mean degree of 4.08 (with $sd = 1.58$) [see Figure \ref{fig:small_comm_vs_large_comm}(a)]. We also find that the neighborhoods of radius four\footnote{We say that a node $j$ belongs to the neighborhood of radius four of node $i$ if the length of the shortest path connecting $i$ and $j$ in the network $y_{\textnormal{ERGM}(k)}$ is smaller than or equal to four.} of PrEP nodes in the large PrEP community of $y_{\textnormal{ERGM}(k)}$ contains a smaller proportion of PrEP nodes ($mean = 26\%$, $sd =0.7\%$) than these neighborhoods of PrEP nodes in the small PrEP communities of $y_{\textnormal{ERGM}(k)}$ ($mean = 29\%$, $sd = 1.3\%$) [see Figure \ref{fig:small_comm_vs_large_comm}(b)].

\section{Simulation results: community structure and syphilis dynamics}
\label{sec:dynamics}

We now use the stochastic syphilis model from Section \ref{sec:syphilis_model_3} to examine potential effects of PrEP usage on syphilis dynamics.  In particular, we use model simulations on the sampled networks $y_{\textnormal{ERGM}(k)}$, $k = 1, \dots, 100$, from the linear ERGM fitted to the NEST data to examine how PrEP usage and community structure impact outbreak probabilities, secondary infections generated, and frequency of reinfections.

\paragraph*{Model parameterization.}
Parameter values for syphilis progression and loss of immunity are based on \cite{tuite2016go} and listed in Table \ref{tab:syphilis_rates_nest}.  Values for the symptom-driven treatment rates $\alpha_1, \alpha_2, \alpha_3, \alpha_4$  are also based on \cite{tuite2016go}.  We calibrate the transmission rate $\beta$ and baseline treatment rate $\alpha$ by comparing long-term prevalence levels from model simulations with empirical prevalence estimates for MSM \cite{tsuboi2021prevalence}.  Specifically, we define the {\it prevalence pseudo-equilibrium} of a model run as the mean percent of nodes with states in $\{I_1, I_2\}$ over times in $[200,240]$ for the run, and compare prevalence pseudo-equilibrium values for the model with empirical prevalence estimates.

We consider two values each for transmission and baseline treatment: low ($\beta=0.05$) and high ($\beta = 0.1$) transmission, and low ($\alpha=0.001$) versus high ($\alpha=0.1$) baseline treatment.  To examine the impact of PrEP, we compare no-PrEP scenarios with $\alpha_{\textnormal{PrEP}} = 0$ for all nodes versus PrEP scenarios where $\alpha_{\textnormal{PrEP}} = 1/1.5$ for PrEP nodes and $\alpha_{\textnormal{PrEP}} = 0$ for non-PrEP nodes.  

\begin{figure}[H]
        \centering
        \subfloat[{\centering Small baseline treatment rate $\alpha = 0.001$}]{\includegraphics[scale = 0.35]{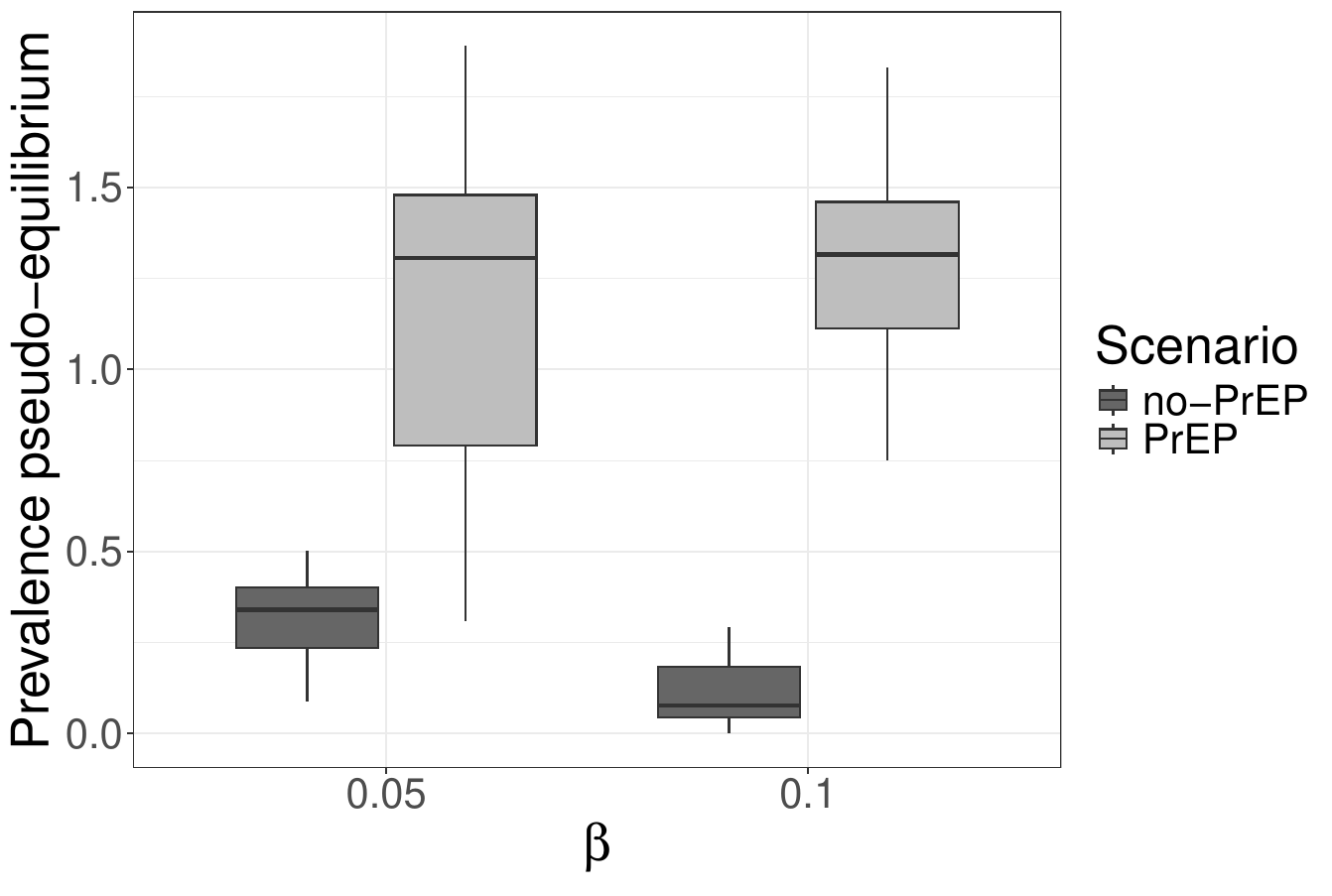}}  \quad
        \subfloat[{\centering Large baseline treatment rate $\alpha = 0.1$}]{\includegraphics[scale=0.345]{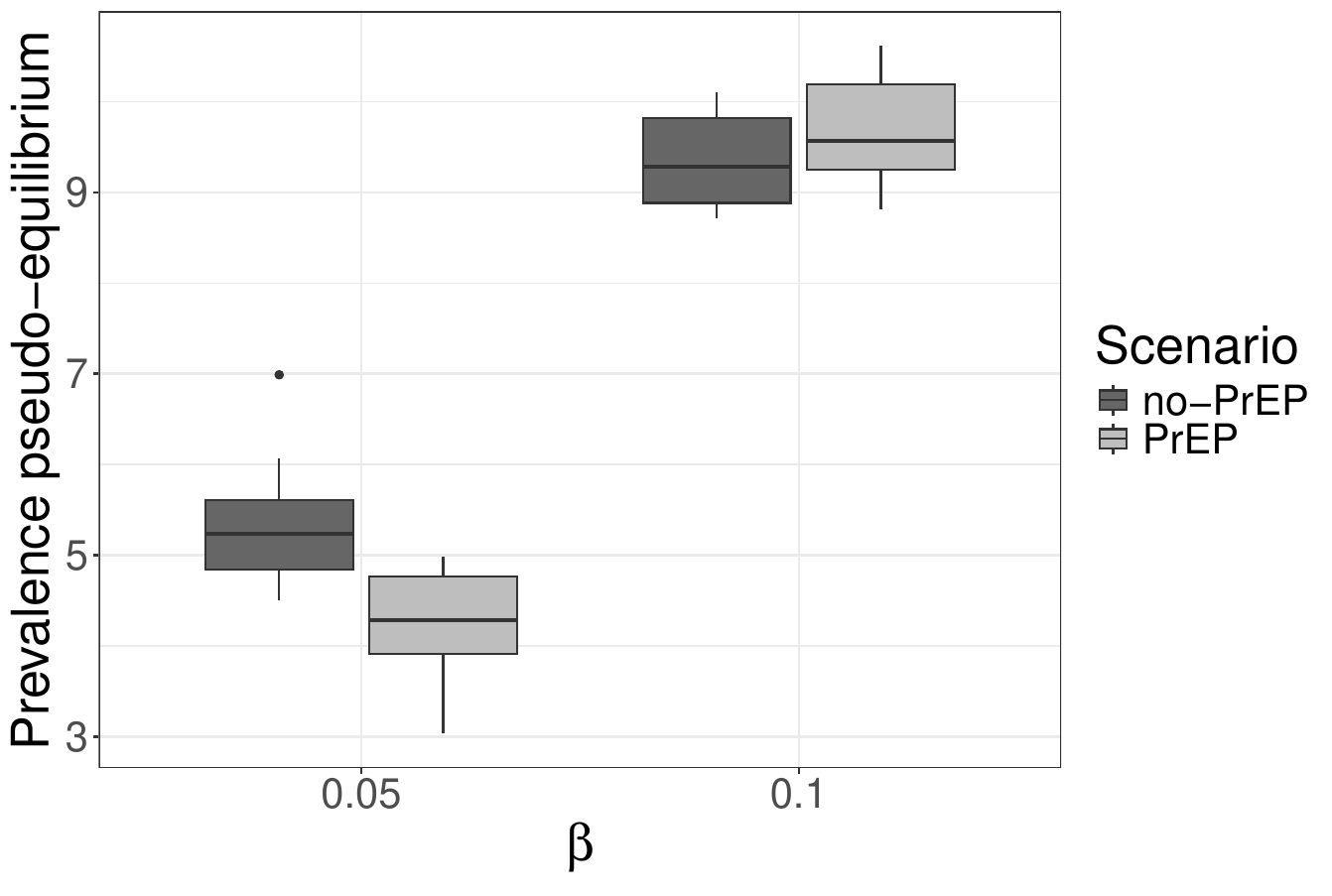}} \\
        \subfloat[{\centering no-PrEP scenario  with $\alpha = 0.1$  and $\beta = 0.05$}]{\includegraphics[scale = 0.5]{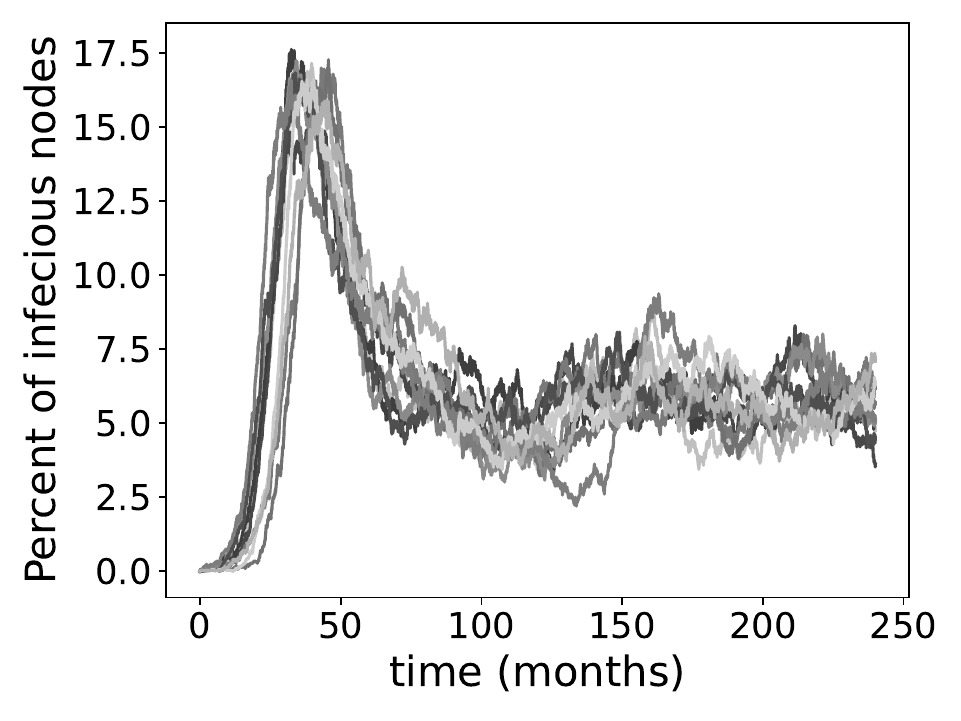}}
        \subfloat[{\centering PrEP scenario with $\alpha = 0.1$ and $\beta = 0.05$}]{\includegraphics[scale=0.5]{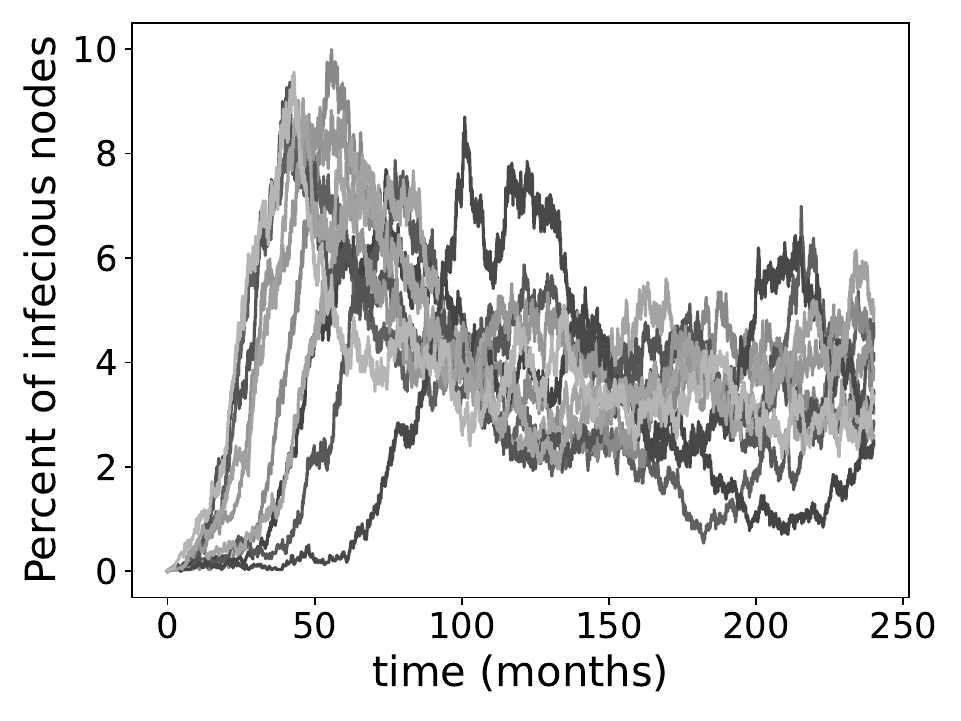}} 
        \caption{Prevalence pseudo-equilibrium distributions and prevalence curves for runs of the syphilis model, where prevalence corresponds to the percentage of nodes with state in $\{I_1,I_2\}$ and prevalence pseudo-equilibrium is defined as mean prevalence over times $[200,240]$ in model runs.          
        (a,b) Distribution of the prevalence pseudo-equilibrium for 5000 model runs with (a) small ($\alpha = 0.001$) and (b) large ($\alpha = 0.1$) baseline treatment rate. (c,d) Prevalence curves 
        of 10 runs of our syphilis model on $y_{{\rm ERGM}(1)}$ with large baseline treatment rate $\alpha = 0.1$ and small transmission rate $\beta = 0.05$ for the (c) no-PrEP scenario and (d) PrEP scenario.    }
        \label{fig:equilibria_long}
\end{figure}

In Figure \ref{fig:equilibria_long}, we show the prevalence pseudo-equilibrium distributions from 5000 total runs (50 runs for each of the networks $y_{\textnormal{ERGM}(k)}$, $k = 1,\ldots, 100$) for the different transmission, baseline treatment, and PrEP scenarios.
For a low baseline treatment rate $\alpha = 0.001$, we observe prevalence pseudo-equilibrium values of less than 1\% (no PrEP) to close to 1.5\% (with PrEP), as shown in Figure \ref{fig:equilibria_long}(a).  These prevalence pseudo-equilibrium values are smaller than estimated syphilis prevalence for MSM \cite{tsuboi2021prevalence}.  Note that for this low baseline treatment rate, including higher treatment rates for PrEP nodes 
leads to larger  prevalence pseudo-equilibrium values than when all nodes are treated as non-PrEP nodes. This result is consistent with the findings of Tuite et al \cite{tuite2016go}, who found that when treatment rates are low, increasing treatment can counter-intuitively lead to increased syphilis transmission due to less people having latent syphilis.  For the high baseline treatment rate $\alpha = 0.1$, we observe prevalence pseudo-equilibrium values of approximately 4\% ($\beta = 0.05$) to more than 9\% ($\beta=0.1$; see Figure \ref{fig:equilibria_long}b). These values are close to the $7.5\%$ global pooled prevalence estimate for MSM by \cite{tsuboi2021prevalence}.  Note that for high baseline treatment rate and low transmission rate ($\alpha = 0.1, \beta = 0.05$) we observe lower prevalence pseudo-equilibrium values in the PrEP scenario versus the no-PrEP scenario.  This finding is again consistent with Tuite et al's model \cite{tuite2016go}, where prevalence decreases with treatment rate provided that the treatment rate is sufficiently high.

\paragraph*{Community structure, PrEP status, and transmission dynamics}
We now examine the impact of community structure and PrEP status on syphilis dynamics.  In particular, we use communities identified by taking PrEP status into account and 
examine how various disease dynamic quantities vary with community membership and PrEP status. The disease dynamics quantities that we examine are long outbreak probability, number of reinfections, and number of secondary infections. We run simulations using either a {\it low basic reproduction number scenario} where $\alpha=0.1$ and $\beta=0.05$, or a {\it high basic reproduction number scenario} where $\alpha=0.001$ and $\beta=0.1$.

We begin by varying initial conditions (ICs) for our syphilis model, running the model starting with a single initial exposed node and all other nodes in the $S$ state, and computing the proportion of simulation runs resulting in long outbreaks as the initial condition is varied. 
Specifically, for each network $y_{\textnormal{ERGM}(k)}$ we  pick uniformly at random 50 PrEP ICs in the large PrEP community, 
50 PrEP ICs in small PrEP communities,
and 50 non-PrEP ICs. For each IC on a given network $y_{\textnormal{ERGM}(k)}$ we run 50 simulations using each of the low basic reproduction number and high basic reproduction number scenarios and aggregate the results for $k=1,\dots,100$. We refer to a simulation run as a \emph{short outbreak} if it has no states in $\{E,I_1,I_2\}$ before 6 months and as a \emph{long outbreak} otherwise.

\begin{figure}[H]
    \centering
    \subfloat[{\centering High basic reproduction number scenario with $\alpha = 0.001$ and $\beta = 0.1$}]{\includegraphics[scale = 0.35]{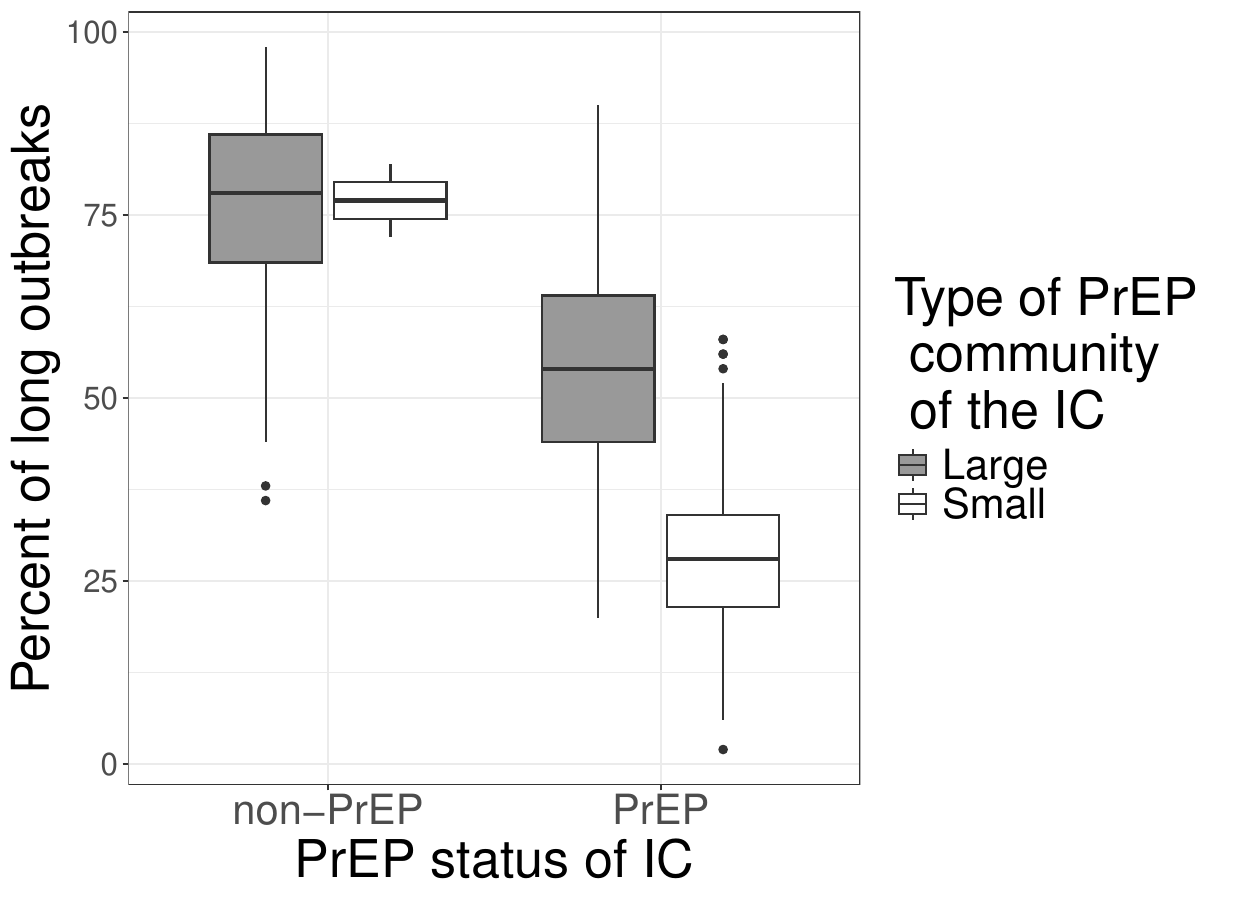}} 
     \subfloat[{\centering Low basic reproduction number scenario with $\alpha = 0.1$ and $\beta = 0.05$}]{\includegraphics[scale = 0.35]{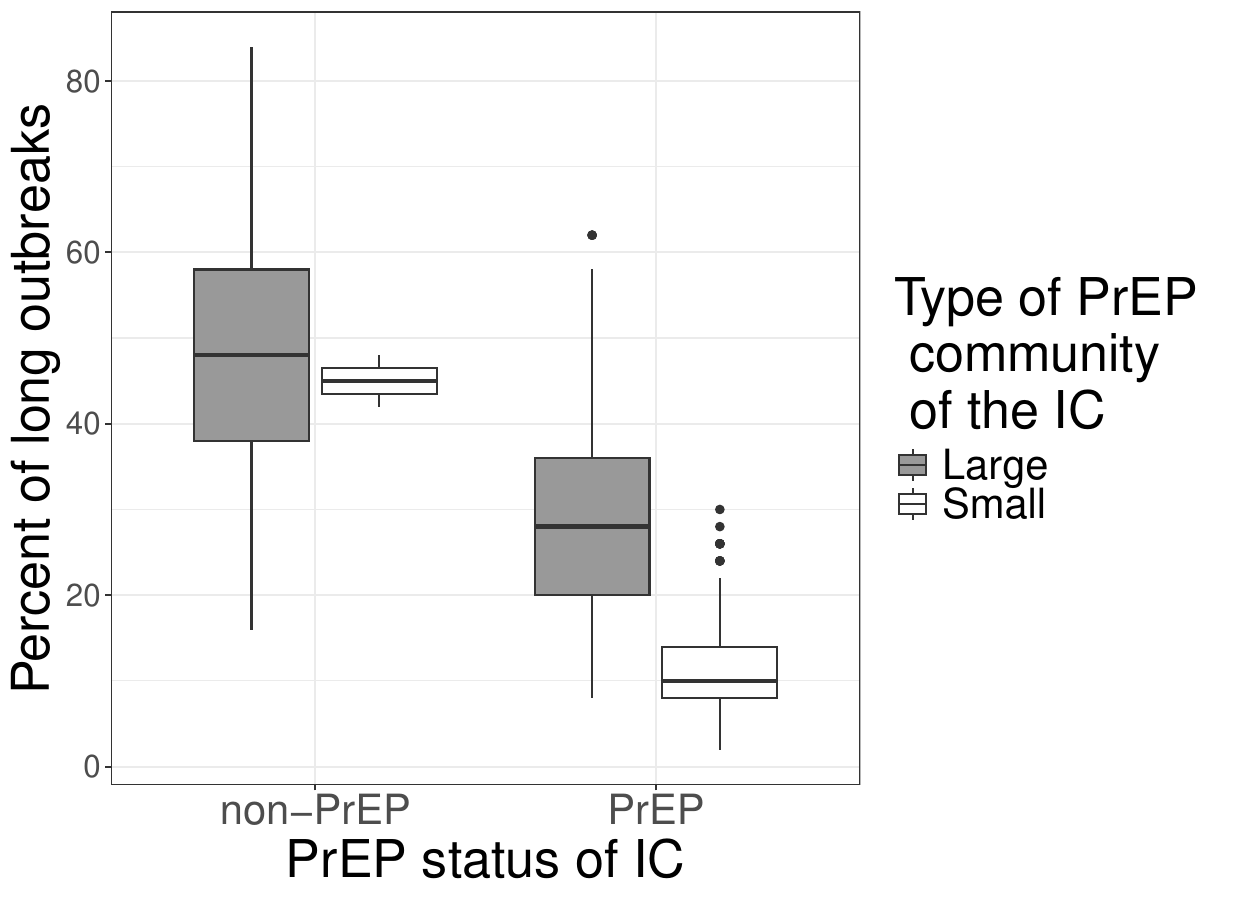}}  
    \caption[]{Distributions of percent of long outbreaks associated with  the ICs in the large no-PrEP community of $y_{{\rm ERGM}(k)}$, for $k\in \{1,\ldots,100\}$. We run 50 simulations per IC and condition to the PrEP status and the type of PrEP community of the IC for $\alpha = 0.001$, $\beta = 0.1$ in (a) and $\alpha = 0.1$, $\beta = 0.05$ in (b). 
    }
    \label{fig:small_comm_short_outbreaks}
\end{figure}

We find that initial infections occurring in non-PrEP nodes lead to a higher percentage of long outbreaks than initial infections occurring in PrEP nodes, regardless of the type of PrEP community for the initially infected node or the basic reproduction number scenario considered.  This result is expected within the small PrEP communities, as PrEP ICs in these small communities have both low degree and a high treatment rate, both of which are associated with decreased transmission.  However, PrEP ICs in the large PrEP community have high degree as well as a high treatment rate, making the net effect on long outbreak probability not obvious.  Our simulation results indicate that for the parameter values considered, increased treatment rates associated with PrEP can compensate for high degree, resulting in lower long outbreak probability for initial infection in PrEP versus non-PrEP nodes within the large PrEP community. 
For example, $53.8\%$ of runs in the high basic reproduction number scenario starting from PrEP nodes in the large PrEP community result in long outbreaks compared with $76.7\%$ of runs starting in non-PrEP nodes in the large PrEP community [see Figure\ref{fig:small_comm_short_outbreaks}(a)]. We observe the same qualitative results in the low basic reproduction number scenario [see Figure \ref{fig:small_comm_short_outbreaks}(b)].

The high treatment rates for PrEP nodes in our model mean that people on PrEP clear syphilis and become susceptible more rapidly than people not on PrEP.  PrEP status may thus influence the number of reinfections a person experiences over a period of time.  This is potentially relevant for disease dynamics on the network as a whole: for example, the 
number of reinfections has been considered as a criterion for the definition of core groups \cite{thomas1996development}. 

\begin{figure}[H]
    \centering
    \subfloat[{\centering High basic reproduction number scenario with $\alpha = 0.001$ and $\beta = 0.1$}]{\includegraphics[scale = 0.35]{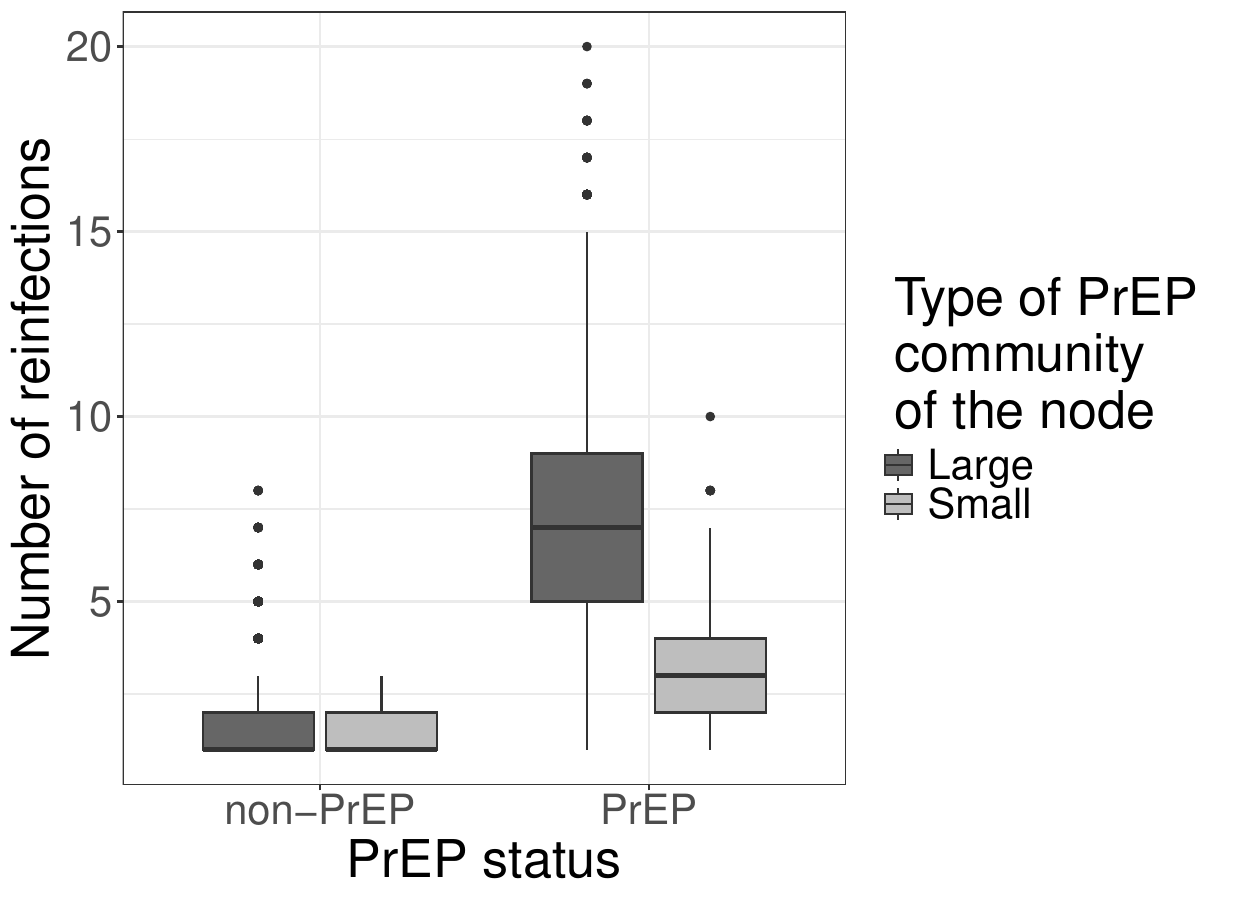}}
    \subfloat[{\centering Low basic reproduction number scenario with $\alpha = 0.1$ and $\beta = 0.05$}]{\includegraphics[scale = 0.35]{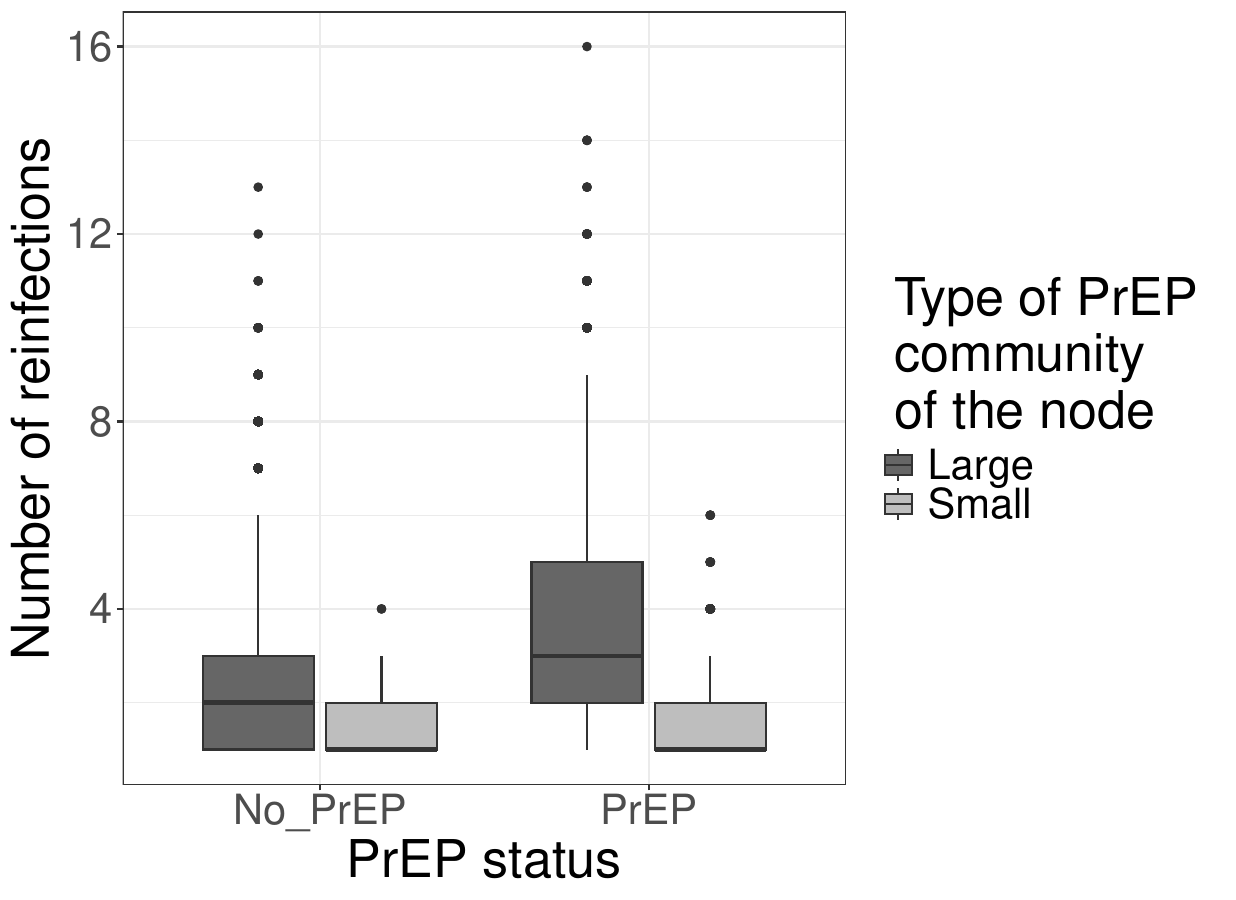}}
    \caption[Reinfections of nodes in the large no-PrEP community]{Distributions of the number of reinfections of the nodes in the large no-PrEP community conditioned on PrEP status and type of PrEP community. These distributions are associated with one run of our syphilis model for 120 months. In (a) we use $\alpha = 0.001$, $\beta = 0.1$. In (b) we use $\alpha = 0.1$, $\beta = 0.05$.
    }
    \label{fig:reinfections}
\end{figure}

In Figure \ref{fig:reinfections}, we compare the number of reinfections for PrEP and non-PrEP nodes in our syphilis model for each of the low and high basic reproduction number scenarios.  For each scenario, we take one model run on each of the networks $y_{\textnormal{ERGM}(k)}$, for $k=1,\ldots,100$. We only select runs where the disease persists for at least 240 months. We then record the number of reinfections that occur during the first 120 months. 

We find that PrEP nodes in the large PrEP community of $y_{\textnormal{ERGM}(k)}$ experience more frequent reinfections compared with PrEP nodes in small PrEP communities or with non-PrEP nodes. For example, for the high basic reproduction number scenario, the mean number of reinfections of PrEP nodes in the large PrEP community of $y_{\textnormal{ERGM}(k)}$ ($mean = 7.18$, $sd = 3.34$) is larger than that of the PrEP nodes in small PrEP communities of $y_{\textnormal{ERGM}(k)}$ ($mean = 2.58$,  $sd = 1.56$) and the non-PrEP nodes of $y_{\textnormal{ERGM}(k)}$ ($mean = 1.69$, $sd = 0.992$) [see Figure \ref{fig:reinfections}(a)]. We observe the same qualitative results in the low basic reproduction number scenario [see Figure \ref{fig:reinfections}(b)].

\begin{figure}[H]
    \centering
    \subfloat[{\centering High basic reproduction number scenario with $\alpha = 0.001$ and $\beta = 0.1$}]{\includegraphics[scale = 0.35]{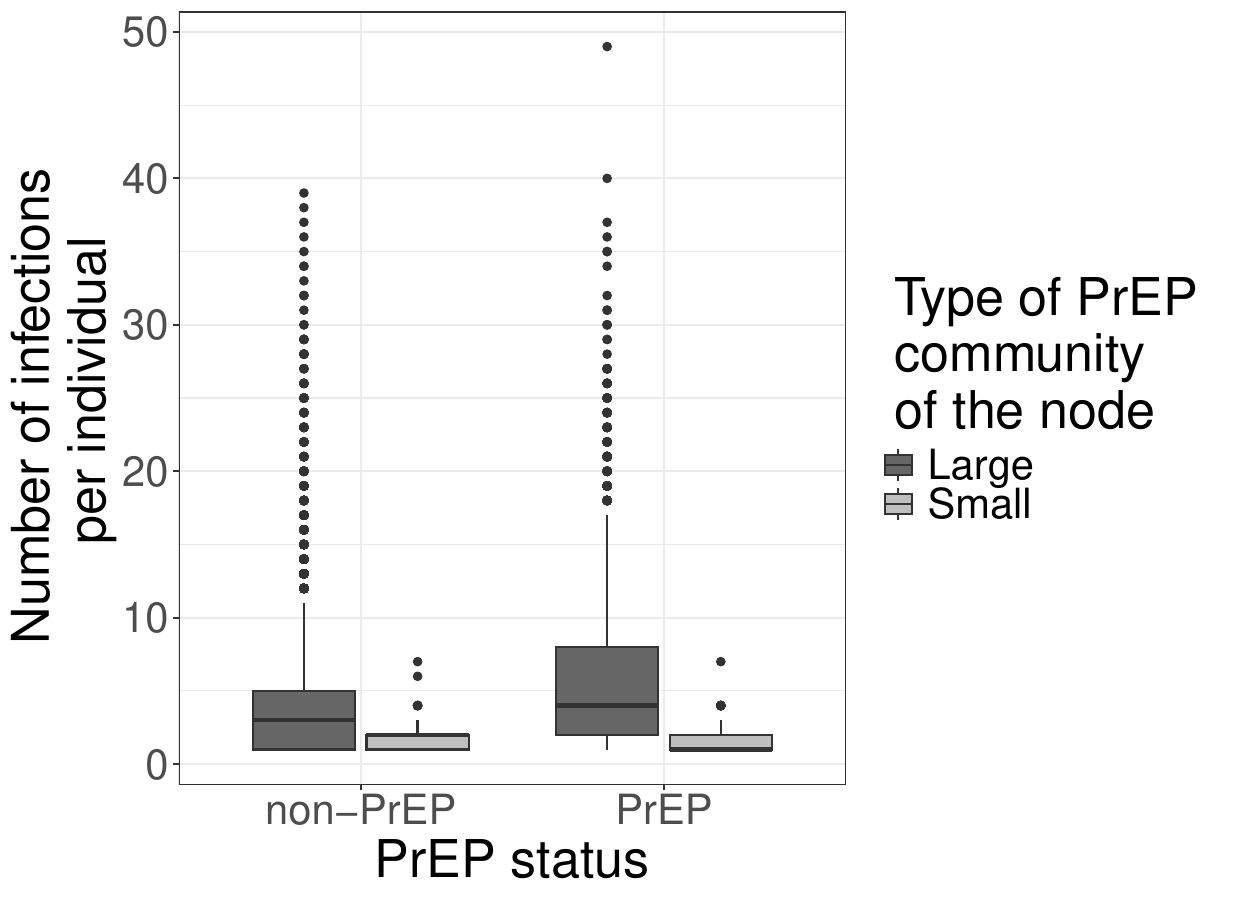}}
    \subfloat[{\centering Low basic reproduction number scenario with $\alpha = 0.1$ and $\beta = 0.05$}]{\includegraphics[scale = 0.35]{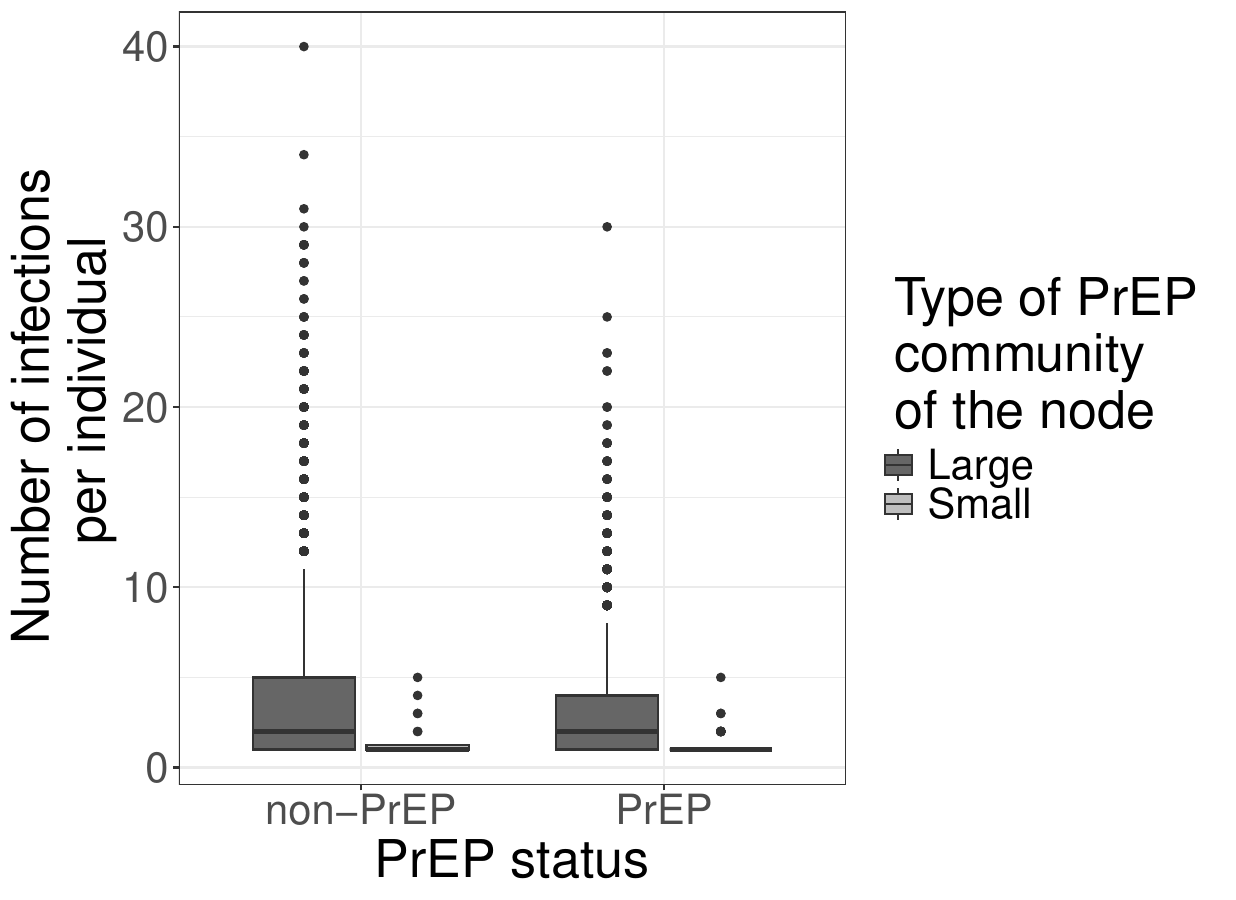}}
    \caption[]{Distributions of the number of infections caused by nodes in the large no-PrEP community conditioned on PrEP status and type of PrEP community. These distributions are associated with one run of our syphilis model for 120 months. In (a) we use $\alpha = 0.001$, $\beta = 0.1$. In (b) we use $\alpha = 0.1$, $\beta = 0.05$.
    }
    \label{fig:reinfectious}
\end{figure}

A greater number of reinfections of the PrEP nodes in the large PrEP community can potentially correspond to a longer time spent infectious and more secondary infections generated.  To evaluate this, we count the number of secondary infections created by nodes according to their PrEP status and PrEP community using the same parameter values and simulation criteria as in the preceding paragraph.  

We find that the number of infections caused by PrEP nodes in the large PrEP community can be relatively high depending on the basic reproduction number scenario. For the high basic reproduction number scenario, we find that the PrEP nodes in the large PrEP community of $y_{\textnormal{ERGM}(k)}$ cause more infections than other nodes. More specifically, the mean number of infections caused by a PrEP node in the large PrEP community of $y_{\textnormal{ERGM}(k)}$ ($mean = 5.58$, $sd = 5.01$) is larger than that of the PrEP nodes in small PrEP communities of $y_{\textnormal{ERGM}(k)}$ ($mean = 1.59$,  $sd = 0.958$) and the non-PrEP nodes of $y_{\textnormal{ERGM}(k)}$ ($mean = 3.71$, $sd = 3.73$) [see Figure \ref{fig:reinfectious}(a)]. This is not the case for the low basic reproduction number scenario, where the number of infections caused by a PrEP node in the large PrEP community ($mean = 3.39$, $sd = 3.07$) is not significantly higher than the number of infections caused by a non-PrEP node in the large PrEP community ($mean = 3.74$, $sd = 3.67$) [see Figure \ref{fig:reinfectious}(b)].

\section{Conclusions and Discussion}\label{sec:discussion_3}
%

%
Our analysis of the NEST data identified two distinct groups of PrEP users: those in a single large community, and those in small communities.  These groups differ in their number of sexual partners, with large community PrEP nodes tending to have higher degree than small community PrEP nodes.  This is relevant for syphilis transmission.  
Our simulation results indicate that these groups of PrEP users play different roles in syphilis transmission.  Small community PrEP users had a small number of sexual partners and consequently fewer secondary transmission events.  These low degree PrEP users are not driving the syphilis epidemic.  

The story regarding large community PrEP nodes is more complex.  PrEP users in this community  {have a large number of contacts}, and create the most secondary infections across community membership and PrEP status in the high reproduction number scenario in our simulations.  In the low transmission scenario, PrEP users and non-PrEP users in the large community create similar numbers of secondary infections. In this case, PrEP and non-PrEP users both contribute to transmission.  
However, the relative difference between the contributions to 
transmission of PrEP and non-PrEP users in the large community depends upon the particular parameters for transmission and treatment.  More research 
here is needed.  

The use of PrEP is correlated to the rise in infections of bacterial sexually transmitted infections {(bacterial STIs)} different than syphilis, such as gonorrhea and chlamydia \cite{kojima2016pre, scott2016sexually}. As in the case of syphilis, PrEP usage is associated with large degree and large treatment rates for gonorrhea and chlamydia. Our methodology is applicable on the NEST data to examine the effect of network connectivity and PrEP usage on the dynamics of these bacterial STIs. More generally, we can apply our methodology to study the effect of network connectivity and treatment on the dynamics of diseases that are not necessarily bacterial STIs where we have access to egocentric data. For example, egocentric data has been used to study HIV transmission \cite{friedman2001social,mustanski2015role, rothenberg2001risk} and non-STIs such as pulmonary infections \cite{miller2021association, potter2011estimating} in small spatial scales. Our methodology can be applied to these cases and help the identification of potential groups driving infection that are informed by treatment or by relevant attributes impacting treatment (e.g., PrEP usage is a relevant attribute impacting syphilis treatment).

If the time scale of treatment is different than the time scale of changes in network connectivity, then modeling disease dynamics on a temporal network is more appropriate than on a static network \cite{holme2012temporal, pastor2015epidemic}. In our methodology, we derived static networks sampled from an ERGM and we examined the effective community structure of these networks using InfoMap for absorbing random walks, which takes static networks as inputs. We also ran our stochastic syphilis model assuming that the treatment rates determined by PrEP usage were constant during the time window of each simulation. However, we observed that some PrEP users stopped reporting PrEP usage in the NEST data. Consequently, a more realistic methodology would require (i) a statistical network generation that accounts for changes in network connectivity and node attributes (e.g., separable exponential random graph models \cite{krivitsky2014separable}) and (ii) a {community detection} algorithm on temporal networks that accounts for treatment changes (e.g., a potential adaptation of InfoMap \cite{edler2017mapping} or modularity maximization {\cite{bazzi2016community}} for temporal networks that accounts for absorption). Nonetheless, the creation of {community detection} algorithms on temporal or multi-layer networks that take into account absorption is still an open problem.

\section*{Code availability}

The code that we use to produce the figures of the manuscript is available at \url{https://github.com/esteban-vargasbernal/NEST_and_absorption}.

\appendix

\section{Appendix}
\label{sec:appendix}

\subsection{Fitted Network} \label{sec:fitted_net}

In this subsection, we present background material on linear ERGMs (Sections \ref{sec:ERGMs_basics} and \ref{sec:ERGM_statistics}), egocentric samples (\ref{sec:egocentric_samples}), and fitting of linear ERGMs (Section \ref{sec:fitting_ERGMs}). In Section \ref{sec:ERGM_NEST}, we present details of the fitted linear ERGM to the NEST data. In Table \ref{tab:ERGM_notation}, we summarize the relevant notation that we use in this subsection. 

\begin{table}[H]
    \centering
    \begin{tabular}{ll}
    \textbf{Symbol}     &  \textbf{Description} \\
    \hline \hline
    $\mathcal{Y}^{(n)}$     &  Undirected graphs with set of nodes $\mathbf{N} =\{1,\ldots, n\}$ \\
    $x_i$ & List of attributes of node $i$ \\
    $x$ & List of node attributes $\left\{x_i \right\}_{i\in \mathbf{N}}$ \\
    $y$ & Network in $\mathcal{Y}^{(n)}$ \\
    $Y$ & Random network in $\mathcal{Y}^{(n)}$ \\
    $\mathbf{y}$ & Set of edges of $y$ \\
    $y + \{i,j\}$ & Network with set of edges  $\mathbf{y}\cup \left\{ \{i,j\}\right\}$ \\
    $y - \{i,j\}$ & Network with set of edges  $\mathbf{y}- \left\{ \{i,j\}\right\}$ \\
    $y_i$ & Set of neighbors of node $i$ in network $y$ \\
    $|y_i|$ & Degree of node $i$ in network $y$ \\
    $g(y)$ & Statistic vector $(g_1(y),\ldots, g_p(y))^{\T}$ defined on $\mathcal{Y}^{(n)}$ \\
    $\mathbb{P}_\theta(Y = y)$ & Probability density function defining a linear ERGM \\
    $\psi(\theta)$ & Normalization factor for $\mathbb{P}_\theta(Y = y)$ \\
    $\theta$ & Coefficient vector of a linear ERGM \\
    $\Delta_{i,j}g(y)$ & Change statistic $g\left(y + \{i,j\}\right) - g\left(y-\{i,j\}\right)$ \\
    $\mathbf{S}$ & Sampled subset of $\mathbf{N}$ \\
    $e_i^{\textnormal{e}}$ & Exogenous information of the ego $i$ \\
    $e_i^{\textnormal{a}}$ & Exogenous information of the alters of the ego $i$ \\
    $e_i$ & Unordered pair $\{e_i^{\textnormal{a}}, e_i^{\textnormal{e}}\}$ \\
    $e_{\mathbf{S}}$ & Egocentric sample $\{e_i\}_{i\in \mathbf{S}}$ \\
    $y_{\textnormal{pop}}$ & Entire network from which we extract $\es$ \\
    $\tilde{n}$ & Scaled-down size with $\tilde{n} \leq n$ \\
    $\tilde{w}_i$ & Observed sample weights \\
    $\var_{\mathbf{S}}\left(\tilde{\theta}\right)$ & Variance of $\tilde{\theta}$ with respect to the sampling of $\mathbf{S}$ \\
    $\mathbb{E}_{\theta}$ & Expectation with respect to $\mathbb{P}_\theta$ \\
    $\var_{\theta}$ & Variance with respect to $\mathbb{P}_\theta$ \\
    $\tilde{g}\left(e_{\mathbf{S}}\right)$ & Estimate of $g(y_{\textnormal{pop}})$ \\
    $l(\theta)$ & Log-likelihood $\theta^{\T}g(y_{\textnormal{pop}}) -\psi(\theta)$ \\
    $l_e(\theta)$ & Log-likelihood $\theta^{\T}\tilde{g}\left(e_{\mathbf{S}}\right) -\psi(\theta)$ \\
    $\theta^*$ & Solution of $\nabla_\theta l(\theta) = \mathbf{0}$ \\
    $\tilde{\theta}$ & Solution of $\nabla_\theta l_e(\theta) = \mathbf{0}$ \\
     \hline 
     \hline
    \end{tabular}
    \caption[Summary of relevant notation for ERGM background]{Summary of relevant notation for our linear ERGM background.
    }
    \label{tab:ERGM_notation}
\end{table}

\subsubsection{Linear ERGMs } \label{sec:ERGMs_basics}

In this subsubsection, we present the definition of a linear ERGM. Let $\mathcal{Y}^{(n)}$ be the set of undirected graphs with set of nodes $\mathbf{N} := \{1,\ldots, n\}$. The space $\mathcal{Y}^{(n)}$ is equipped with an unordered list of \emph{exogenous information} $x=\left\{x_i\right\}_{i\in \mathbf{N}}$, where $x_i$ is a list of attributes of the node $i$ (e.g., age, race, etc). For $y\in \mathcal{Y}^{(n)}$, let $\mathbf{y}$ denote the set of edges of $y$, and let $|\mathbf{y}|$ denote the total number of edges of $y$. We define $y_{i,j}$ by $y_{i,j} = 1$ if $\{i,j\} \in \mathbf{y}$ and $y_{i,j}=0$ otherwise. Let $y_i$ denote the set of neighbors of the node $i$. Let $y+\{i,j\}$ denote the network with set of edges $\mathbf{y}\cup \left\{\{i,j\}\right\}$, and let $y-\{i,j\}$ denote the network with set of edges $\mathbf{y}-\left\{\{i,j\}\right\}$. In Definition \ref{def:ergm}, we present the probability density functions that define linear ERGMs.

\begin{definition}{\bf (Linear exponential random graph models)}\label{def:ergm}
    A \emph{linear exponential random graph model (linear ERGM)} is a probability density function $\mathbb{P}_\theta(Y=y)$ defined on $\mathcal{Y}^{(n)}$ by
\begin{equation}\label{eqn:Pr_ERGM}
    \mathbb{P}_{\theta}(Y=y) := \exp\left\{\theta^{\T} g(y) - \psi(\theta)\right\}, \quad y\in \mathcal{Y}^{(n)} \,,
\end{equation}
where $g(y) = \left(g_1(y), \ldots, g_p(y)\right)^{\T}$ is a statistic vector, and 
$$ \exp \left\{ \psi(\theta) \right\} := \sum_{z\in \mathcal{Y}^{(n)}} \exp\left(\theta^{\T} g(z)\right)\,.$$
\end{definition}

\subsubsection{Some ERGM statistics} \label{sec:ERGM_statistics}

In Table \ref{tab:stats_fitted} of Section \ref{sec:the_network}, we presented the statistics that we use in the fitted linear ERGM. These statistics are classified as propensity, homophily, and monogamy statistics. In Table \ref{tab:sufficient_stats}, we present the definitions of propensity, homophily, and monogamy of a generic subset $A$ of the set of nodes.  

\begin{table}[H]
    \centering
    \begin{tabular}{lll}
    \textbf{Statistic} $\boldsymbol{g_k(y,x)}$ & \textbf{Definition}   \\
    \hline \hline
    \textbf{Propensity of group} $A$: number of times & $\sum_{\{i,j\} \in \mathbf{y}} (1_{i\in A}+1_{j\in A})$ \\
    a node in $A\subseteq \{1,\ldots , n\}$ appears in an edge &  \\
    \textbf{Homophily of group} $A$: number of edges & $\sum_{\{i,j\} \in \mathbf{y}} 1_{i\in A} 1_{j\in A}$\\
     with nodes inside $A\subseteq \{1,\ldots,n\}$ &  \\
    \textbf{Monogamy of group} $A$: number of nodes & $\sum_{i\in \mathbf{N}} 1_{|y_i|=1}1_{i\in A}$\\
     with degree $1$ in $A \subseteq \{1,\ldots,n\}$ & \\
    \hline
    \hline
    \end{tabular}
    \caption[Statistics used in the ERGM fitted to the NEST data]{Types of statistics that we use in Section \ref{sec:the_network}. The expressions $1_{\textnormal{``property''}}$ on the second column are defined by $1_{\textnormal{``property''}} = 1$ if $\textnormal{``property''}$ is true and $1_{\textnormal{``property''}} = 0$ otherwise. }
    \label{tab:sufficient_stats}
\end{table}

The size $n$ of the MSM population is large (and so is the space $\mathcal{Y}^{(n)}$), making the fitting of our linear ERGM computationally challenging. Consequently, we assume that the population size is $\tilde{n}$, with $\tilde{n} \leq n$ (we refer to $\tilde{n}$ as \emph{scaled-down size}). This assumption can affect the mean degree distribution of the drawn networks from the fitted linear ERGM \cite{krivitsky2011adjusting}. This is not a desirable property for networks where people make connections as function of local information. To mitigate the effect of the scale-down size on the mean degree of networks drawn from linear ERGMs\footnote{For a linear ERGM $\mathbb{P}_\theta$ with fixed coefficient vector $\theta$, the expected degree of a drawn network from the distribution $\mathbb{P}_\theta$ depends on the size of the network. The scaled-down statistic and the scaled-down coefficient mitigate the effect of the scale-down size on the expected degree of networks drawn from the distribution $\mathbb{P}_\theta$ \cite{krivitsky2011adjusting}.}, an additional statistic with fixed coefficient is used \cite{krivitsky2011adjusting}. More specifically, the statistic $|\mathbf{y}|$ (that we call \emph{scaled-down statistic}) is included in the model that we fit in Section \ref{sec:fitted_to_NEST}. This scaled-down statistic has a fixed coefficient $-\log(\tilde{n})$ (that we call \emph{scaled-down coefficient}), where ``$\log$'' denotes the natural logarithm function. Consequently, the fitted linear ERGM satisfies 
\begin{equation*}
    \mathbb{P}_{\theta}\left(Y=y\right) \propto \exp\left(-\log(\tilde{n})|\mathbf{y}| + \tilde{\theta}^{\T}g(y)\right) \,,
\end{equation*}
where $g(y) = \left(g_1(y),\ldots, g_{20}(y)\right)^{\T}$ is the statistic vector introduced in Section \ref{sec:fitted_to_NEST}.

\subsubsection{Egocentric samples}\label{sec:egocentric_samples}

In this subsubsection, we introduce the concepts of egocentric samples (such as the egocentric sample from the NEST data) and sample weights that are related to the fitted linear ERGM in Section \ref{sec:the_network}. For a node $i$ of a network $y \in \mathcal{Y}^{(n)}$ with exogenous information $x$, let $e_i^{\textnormal{e}} :=  x_i$ denote the attributes of $i$ and let $e_i^{\textnormal{a}}$ denote the unordered list $\{x_j\}_{j\in y_i}$ of attributes of the neighbors of $i$. We define the \emph{egocentric view of $y$ from the ego $i$} as the unordered pair $\{e_i^{\textnormal{e}}, e_i^{\textnormal{a}}\}$ and we denote it by $e_i$. We define \emph{an egocentric sample of $y$} as the unordered list $\{e_i\}_{i\in \mathbf{S}}$, for some $\mathbf{S} \subseteq \mathbf{N}$, and we denote it by $e_{\textbf{S}}$. 

An \emph{egocentric sampled network} consists of a sample $\mathbf{S}$ of $\mathbf{N}$ (the nodes of $\mathbf{S}$ are called \emph{egos}, and these nodes represent the respondents of the NEST study in this work), and a list of nodes (which are called \emph{alters}) with whom the egos had a relationship. The egos provide exogenous information $e_i^{\textnormal{e}}$ of themselves and exogenous information $e_i^{\textnormal{a}}$ of their alters. In this work, we use \emph{minimal egocentric networks}, where the alters cannot be uniquely identified (one person could be reported as alter by two different egos, having two different alter identities in the data), and the attributes of the exogenous information for the egos and the alters are the same. An observed egocentric sampled network then leads to an egocentric sample $e_{\textbf{S}}$ of the network $y_{\textnormal{pop}}$ associated with the entire population. 

To compensate for biases due to sampling, we consider the sample weights $w_i \propto \left[ \mathbb{P}\left(i\in \mathbf{S}\right) \right]^{-1}$ that adjust the distribution of attributes that might have been over-sampled or under-sampled in $\mathbf{S}$. We treat $w_i$ as a random variable determined by the random sample $\mathbf{S}$. Let $\tilde{w}_i$ denote the sample weights associated with a specific egocentric sampled network that is observed.

\subsubsection{Fitting ERGMs}\label{sec:fitting_ERGMs}

In this subsubsection, we present the algorithm Pseudo Maximum Likelihood Estimation (PMLE) that is used to obtain an estimate $\tilde{\theta}$ of the coefficient vector of a linear ERGM from an observed egocentric sample $e_{\mathbf{S}}$ and sample weights $\tilde{w}_i$. We follow the presentation of PMLE in \cite{krivitsky2017inference}. 

Let $\mathbb{P}_\theta$ be a linear ERGM with static vector $g(y)$, where we can write the statistics $g_k$ as

\begin{align}\label{eqn:ego_stat}
  g_k(y) = \sum_{\{i,j\} \in \mathbf{y}} f_k(x_i,x_j) = \sum_{ i\in\mathbf{N}} h_k(e_i) \,,  
\end{align}
with $h_k(e_i) := \frac{1}{2} \sum_{j \in y_i}f_k(x_i, x_j)$. Notice that each statistics in Table \ref{tab:sufficient_stats} can be written in the form of one of the two summations in (\ref{eqn:ego_stat}).

We have that the log-likelihood of $\mathbb{P}_\theta$ given a network $y_{\textnormal{pop}}$ is 
\begin{align*}
    l(\theta) = \theta^{\T}g\left(y_{\textnormal{pop}}\right) - \psi(\theta) \,.
\end{align*} 
Given an egocentric sample $e_{\mathbf{S}}$ of $y_{\textnormal{pop}}$, we need to estimate a coefficient vector that maximizes the log-likelihood $l(\theta)$ over the space of undirected networks $\mathcal{Y}^{(\tilde{n})}$. If the network $y_{\textnormal{pop}}$ is known, then a coefficient vector $\theta^*$ that maximizes $l(\theta)$ satisfies   

\begin{align}\label{eqn:gradl}
    \nabla_\theta l(\theta^*) = g(y_{\textnormal{pop}})-\mathbb{E}_{\theta^*} \left(g(Y) \right)  = \mathbf{0} \,, 
\end{align}
where $\mathbb{E}_\theta \left( g(Y) \right)= \sum_y g(y) \mathbb{P}_\theta(Y = y)$. However, the network $y_{\textnormal{pop}}$ is unknown in our study and we only know the egocentric sample $\es$ and sample weights $\tilde{w}_i$. We then estimate $g(y_{\textnormal{pop}})$ by 

\begin{align}\label{eqn:g_tilde}
\tilde{g}(\es)  = \sum_{i \in \mathbf{S}} \left(\frac{\tilde{n} \tilde{w}_i}{\sum_{j \in \mathbf{S}}\tilde{w_j}} \right) h(e_i)\,,
\end{align}
where $h(e_i) = (h_1(e_i),\ldots, h_k(e_i))^{\T}$.

If we define 
\begin{align*}
l_e(\theta) := \theta^{\T} \tilde{g}(\es) - \psi(\theta) \, ,
\end{align*}
we then  find an estimate $\tilde{\theta}$ of $\theta^*$ such that

\begin{align}\label{eqn:gradle}
\nabla_\theta l_e(\tilde{\theta}) = \tilde{g}(\es) - \mathbb{E}_{\tilde{\theta}} \left(g(Y) \right) = \mathbf{0} \,.
\end{align}
The solution $\tilde{\theta}$ of (\ref{eqn:gradle}) can be found using Monte Carlo Maximum Likelihood Estimation (MCMLE) \cite{hunter2006inference}. 

Let $\var_{\mathbf{S}}\left(\tilde{\theta}\right)$ be the variance of $\tilde{\theta}$ with respect to the randomness of the sample $\mathbf{S}$. We have that $\tilde{\theta}$ is a consistent, asymptotically normal estimator of $\theta^*$ [where $\theta^*$ solves (\ref{eqn:gradl})], and 
\begin{align}\label{eqn:CLTtheta}
    \tilde{\theta}- \theta^* \xrightarrow{d} \textnormal{MVN}_p\left(\mathbf{0}, \var_{\mathbf{S}}\left(\tilde{\theta}\right) \right)\,
\end{align}
as $|\mathbf{S}|\rightarrow \tilde{n}$ \cite{binder1983variances}, where $\textnormal{NMV}_p$ denotes a $p$-dimensional multivariate normal distribution and the symbol ``$\xrightarrow{d}$'' denotes convergence in distribution. The method MCMLE also gives an estimate  $\widetilde{\var}_{\tilde{\theta}}\left(g(Y)\right)$ of $\var_{\tilde{\theta}}\left(g(Y)\right)$, which is a variance with respect to the distribution $\mathbb{P}_{\tilde{\theta}}$. 
We can use this estimate and the sample weights $\tilde{w}_i$ to obtain an estimate $\widetilde{\var}_{\mathbf{S}}\left(\tilde{\theta}\right)$ of $\var_{\mathbf{S}}\left(\tilde{\theta}\right)$ (see equation (4.5) of \cite{krivitsky2017inference}). The estimate $\widetilde{\var}_{\mathbf{S}}\left(\tilde{\theta}\right)$ is used to test if a specific coefficient is significantly different from zero. If a coefficient is significantly different from zero, then the corresponding statistic impacts the probability of edge occurrence [see equation (\ref{eqn:edge_prob})]. In Algorithm \ref{alg:PMLE}, we summarize the steps of PMLE to find $\tilde{\theta}$ and $\widetilde{\var}_{\mathbf{S}}\left(\tilde{\theta}\right)$. 

\begin{algorithm}[H]
\caption{Pseudo Maximum Likelihood Estimation.}
\label{alg:PMLE}
\begin{algorithmic}[1]
 \renewcommand{\algorithmicrequire}{Input:}
 \renewcommand{\algorithmicensure}{Output:}
 \REQUIRE An egocentric sample $e_{\mathbf{S}}$ of a network $y_{\textnormal{pop}}$ with size $\tilde{n}$, sample weights $\tilde{w}_i$, for $i\in \mathbf{S}$, and statistics $g_k(y)$, for $k=1,\ldots,p$. 
 \ENSURE   The estimate $\tilde{\theta}$ that satisfies (\ref{eqn:gradle}) and the estimated variance $\widetilde{\var}_{\mathbf{S}}\left(\tilde{\theta}\right)$.
	\STATE Compute $\tilde{g}(\es)$ using (\ref{eqn:g_tilde}).
	\STATE Use Monte Carlo Maximum Likelihood Estimation to obtain $\tilde{\theta}$ such that 
	\begin{align*}
    \tilde{g}(\es) - \mathbb{E}_{\tilde{\theta}} \left(g(Y) \right) = \mathbf{0} \,.
    \end{align*}
	\STATE From the Monte Carlo Maximum Likelihood Estimation, obtain $\widetilde{\var}_{\tilde{\theta}}\left(g(Y)\right)$.
	\STATE Use $\tilde{w}_i$, for $i \in \mathbf{S}$, and $\widetilde{\var}_{\tilde{\theta}}\left(g(Y)\right)$ to obtain the estimate $\widetilde{\var}_{\mathbf{S}}\left(\tilde{\theta}\right)$ (using equation (4.5) of \cite{krivitsky2017inference}).
\end{algorithmic}
\end{algorithm}

\subsubsection{Fitted linear ERGM to the NEST data}\label{sec:ERGM_NEST}

In this subsubsection, we present the ERGM that we fit to the NEST egocentric sample $\es$ presented in Section \ref{sec:the_network}. We consider a linear ERGM that uses the statistics $g_1,\ldots,g_{20}$ in Table \ref{tab:stats_fitted} of Section \ref{sec:fitted_to_NEST} and a scale-down statistic $g_0$ with scale-down size $\tilde{n} = 3000$ (so $\tilde{n}/|\mathbf{S}| =  12.7$, which is a similar ratio as in the ERGM fitted in \cite{krivitsky2017inference} for The National Health and Social Life Survey data\footnote{Krivitsky and Morris \cite{krivitsky2017inference} fitted linear ERGMs to the egocentric sampled network of the National Health and Social Life Survey (NHSLS) with the goal of understanding persistent racial disparities in HIV prevalence in United States.}). We obtain sample weights $\tilde{w}_i$ to match the age and race distributions of the census of Franklin county. This procedure follows the ranking algorithm in \cite{debell2009computing}. We use the R package \emph{anesrake} \cite{pasek2018package} to obtain the sample weights $\tilde{w}_i$. We use the R package \emph{ergm.ego} \cite{krivitsky2021ergm} for fitting the linear ERGM with statistics in $g^{\textnormal{model}} := (g_0,\ldots ,g_{20})^{\T}$ to $e_{\mathbf{S}}$ and $\tilde{w}_i$. This R package rounds the size of the networks generated for the PMLE to an integer using the sample weights $\tilde{w}_i$ (see details on this rounding in \cite{krivitsky2017inference}). For our fitted linear ERGM, $\tilde{n}$ is rounded to 2978. This rounded size is also the size of the networks $\tilde{y}_{(1)}, \ldots, \tilde{y}_{(100)}$ that we draw from the distribution $\mathbb{P}_{\tilde{\theta}}$ in Section \ref{sec:the_network}, where $\tilde{\theta}$ is the coefficient vector that is fitted by Algorithm \ref{alg:PMLE}.

In Table \ref{tab:fitted_ergm}, we give the relevant information of the fitted linear ERGM to the NEST data. In the first column of Table \ref{tab:fitted_ergm}, we give the statistics of our fitted linear ERGM. In the second column of Table \ref{tab:fitted_ergm}, we give the estimates of the coefficients $\theta_i$ that we obtain from Algorithm \ref{alg:PMLE}. In the third column of Table \ref{tab:fitted_ergm}, we give the errors $\sqrt{\widetilde{\var}_{\mathbf{S}}\left(\theta_i\right)}$ that we obtain from Algorithm \ref{alg:PMLE}. In the fourth column of Table \ref{tab:fitted_ergm}, we give the z-values $\theta_i\left/\sqrt{\widetilde{\var}_{\mathbf{S}}\left(\theta_i\right)}\right.$. In the fifth column of Table \ref{tab:fitted_ergm}, we give the p-values associated with the null hypothesis $H_0: \theta_i = 0$.

\begin{table}
\centering
\resizebox{\textwidth}{!}{
\begin{tabular}{lllll}

\textbf{Statistic} $\boldsymbol{g_i}$ & \textbf{Coefficient} $\boldsymbol{\theta_i}$ & \textbf{Std. error} &  \textbf{z-value} & $\boldsymbol{\mathbb{P}(>z)}$ \\
\hline \hline

$g_0:$ Scaled-down statistic   &  $\theta_0 = $ -7.99901 & 0.00000 &-Inf  & < 1e-04 *** \\

\hline 
$g_1:$ Propensity of age group 1 &  $\theta_1 = $ 0.61107 & 0.29100 &    2.100 & 0.03574 * \\ 
$g_2:$ Propensity of age group 2 &  $\theta_2 = $ 0.91969 & 0.25879 &     3.554 & 0.00038 *** \\
$g_3:$ Propensity of age group 3 &  $\theta_3 = $ 0.96084 & 0.21096 &    4.555 & < 1e-04 *** \\

\hline 
$g_4:$ Propensity of white race &   $\theta_4 = $ 0.55611 & 0.44799  &  1.241 &  0.21448 \\    
$g_5:$ Propensity of black race &   $\theta_5 = $ -0.15265 & 0.34194 &    -0.446 &  0.65529  \\

\hline
$g_6:$ Propensity of PrEP group     & $\theta_6 = $ -0.73628 &  0.80615 &   -0.913 &  0.36107 \\    

\hline
$g_7:$ Propensity of sex-trade group & $\theta_7 = $ -0.09369 & 0.36699 &  -0.255 &  0.79849  \\  

\hline
$g_8:$ Propensity of group-sex group & $\theta_8 = $ 0.78618  & 0.13697 &  5.740 &  < 1e-04 *** \\

\hline
$g_9:$ Homophily of age group 1 &  $\theta_9 = $ 2.11656  & 0.26252 & 8.062 & < 1e-04 *** \\
$g_{10}:$ Homophily of age group 2 &  $\theta_{10} = $ 0.61010 & 0.20441 &   2.985 &  0.00284 ** \\
$g_{11}:$ Homophily of age group 3 &  $\theta_{11} = $ -0.13726 &  0.21228 &  -0.647 & 0.51787    \\
$g_{12}:$ Homophily of age group 4 &   $\theta_{12} = $ 1.25233 & 0.26804 &    4.672 &  < 1e-04 *** \\

\hline 
$g_{13}:$ Homophily of white race  &    $\theta_{13} = $ 0.19513 & 0.35089 &    0.556 &  0.57814 \\ 
$g_{14}:$ Homophily of black race &    $\theta_{14} = $ 2.27063 & 0.38016 &   5.973 &  < 1e-04 *** \\
$g_{15}:$ Homophily of other race  &   $\theta_{15} = $ -2.91303 & 1.12805 &   -2.582 &  0.00981 ** \\

\hline
$g_{16}:$ Homophily of no-PrEP group    &   $\theta_{16} = $ -0.91144 &  0.81365 &  -1.120 & 0.26263  \\  
$g_{17}:$ Homophily of PrEP group      &    $\theta_{17} = $ 0.61785 & 0.82057 &    0.753 & 0.45148 \\    

\hline 
$g_{18}:$ Monogamy of white race       &    $\theta_{18} = $ 2.91975 &  1.17451 &  2.486 & 0.01292 * \\ 
$g_{19}:$ Monogamy of black race     &    $\theta_{19} = $ 1.90546 & 1.08326 &   1.759  &  0.07858 .  \\
$g_{20}:$ Monogamy of other race    &   $\theta_{20} = $ -0.14570 &  1.42531 &   -0.102  &  0.91858  \\  

\hline 
\hline
\end{tabular}
}
\caption[Fitted linear ERGM to the NEST data]{Fitted linear ERGM to the NEST data.}
\label{tab:fitted_ergm}
\end{table}

\subsection{InfoMap for absorbing random walks}\label{appendix:infomap}

In this subsection, we present the adaptation of InfoMap to absorbing random walks that we use in Section \ref{sec:community}. Let $z$ be a directed and weighted network with adjacency matrix $A = (a_{ij})_{i,j\in\{1,\ldots,n\}}$ (i.e., $a_{ij} >0$ when the network has a directed edge $(j,i)$, and $a_{ij} = 0$ otherwise). Let $W:= \diag\{\omega_1,\ldots,\omega_n\}$, where $\omega_j := \sum_i a_{ij} \neq 0$ for $j  \in \{1,\ldots,n\}$. We can then define a Markov chain on $z$ with transition probability matrix $P := AW^{-1}$. 

In the following definition, we describe the standard map function evaluated at a partition of the set of nodes of $z$ when $P$ is a regular matrix (i.e., all the entries of some power of $P$ are positive) \cite{rosvall2008maps}. (See \cite{bohlin2014community} for the definition of $L(M)$ when $P$ is not regular.) 

\begin{definition}{\bf (Standard map function)}\label{def:Map1}
Let $\pi = (\pi_1,\ldots,\pi_n)^{\T}$ be the stationary distribution associated with $P$. For a partition $M= \{M_1,\ldots,M_m\}$ of the set of nodes $\{1,\ldots,n\}$ of $z$, let  $q_{i\curvearrowright} := \sum_{j \in M_i, k \notin M_i} \pi_j p_{kj}$, $q_{i\curvearrowleft} := \sum_{k \in M_i, j \notin M_i} \pi_j p_{kj}$, $q_\curvearrowleft := \sum_{i \in \{1,\ldots,m\}} q_{i\curvearrowleft}$, and $p_\circlearrowright^i := q_{i\curvearrowright } + \sum_{j\in M_i} \pi_j$. 

Let $\mathcal{H}(\mathcal{Q})$ be the entropy\footnote{Recall that the entropy of a distribution with strictly positive probabilities $p_1,\ldots,p_r$ is $\mathcal{H}(\{p_1,\ldots, p_r\}) := -\sum_i p_i\log_2(p_i)$.} of the distribution $\mathcal{Q}$, where the probabilities of $Q$ are 
\begin{equation*} 	
	q_{1\curvearrowleft}/q_\curvearrowleft, \ldots, q_{m\curvearrowleft}/q_\curvearrowleft \, . 
\end{equation*}	

Let $\mathcal{H}(\mathcal{P}^i)$ be the entropy of the distribution $\mathcal{P}^i$, where the probabilities of $\mathcal{P}^i$ are
\begin{equation*}
 q_{i \curvearrowright}/p_\circlearrowright^i, \pi_{k_1}/p_\circlearrowright^i, \ldots, \pi_{k_i}/p_\circlearrowright^i  \,,
\end{equation*}	
where $M_i = \{k_1,\ldots,k_i\}$. The \emph{standard map function that is associated with $P$} is 
\begin{equation*}
	L(M) :=  q_\curvearrowleft \mathcal{H}(\mathcal{Q}) + \sum_{i=1}^m p_\circlearrowright ^i \mathcal{H}(\mathcal{P}^i)\,.
\end{equation*}	
\end{definition}
Minimizing $L(M)$ gives a partition with dense connections within the elements of $M$ and sparse connections between the elements of $M$. \emph{Standard InfoMap} with input $P$ is an algorithm to minimize $L(M)$ \cite{rosvall2009map,rosvall2008maps}.   

We now describe the adaptation of InfoMap to absorbing random walks \cite{vargas2022infomap} that we use in this paper. Let $y$ be a connected network in $\mathcal{Y}^{(n)}$ with adjacency matrix $A$ such that $AW^{-1}$ is regular. Let $\delta_1, \ldots, \delta_n$ be positive scalars, which are called \emph{node-absorption rates}. Let us define a directed and weighted network $z$ with set of nodes $\{1, \ldots, n+1\}$ such that: i) $z$ has edges $(i,j)$ and $(j,i)$ with weights equal to one if and only if $\{i,j\}$ is in $\mathbf{y}$, ii) $z$ has edges $(i,n+1)$ with weights $\delta_i$ for $i=1,\ldots,n$, and iii) $z$ has no edges coming out of $n+1$. We can then define an absorbing Markov chain on $z$ with absorbing state $n+1$, and transition probability matrix 

\begin{align*}
    \tilde{P} = \begin{pmatrix}
                    A(W+D_\delta)^{-1} & \mathbf{0} \\
                    \vec{\delta}^{\T}(W+D_\delta)^{-1}& 1
                \end{pmatrix} \, ,
\end{align*}
where $\vec{\delta} := (\delta_1, \ldots, \delta_n)^{\T}$ is called \emph{node-absorption-rate vector} and $D_\delta := \diag \left\{\vec{\delta}\right\}$. To examine the effect of $\vec{\delta}$ on the community structure of the network $y$, we use standard InfoMap for the transition-probability matrix
\begin{equation}\label{eqn:Ptexp}
	 P(\vec{\delta},t) := e^{-t \mathcal{L}} \, ,    
\end{equation}
where $t$ is a positive constant (which is called \emph{Markov time}), and $\mathcal{L} := (W-A)D^{-1}_\delta$. The matrix $\mathcal{L}$ is the unnormalized graph Laplacian of the absorption-scaled graph with adjacency matrix $AD_\delta^{-1}$. This absorption-scaled graph is closely related to the absorbing Markov chain with transition-probability matrix $\tilde{P}$ \cite{jacobsen2018generalized, vargas2022infomap}. The matrix $P(\vec{\delta},t)$ is the transition-probability matrix of a continuous-time Markov chain with time steps of length $t$ and infinitesimal generator $-\mathcal{L}$. 
In Algorithm \ref{alg:adapt_info}, we summarize InfoMap for absorbing random walks.


\begin{algorithm}
\caption{InfoMap for absorbing {random walks}.}
\label{alg:adapt_info}
\begin{algorithmic}[1]
 \renewcommand{\algorithmicrequire}{Input:}
 \renewcommand{\algorithmicensure}{Output:}
 \REQUIRE An adjacency matrix $A=(a_{ij})_{i,j\in \{1,\ldots,n\} }$ {of a connected graph} in $\mathcal{Y}^{(n)}$ such that $AW^{-1}$ is regular, and a node-absorption-rate vector $\vec{\delta}=(\delta_1,\ldots,\delta_n)^{\T}$ with positive entries.
 \ENSURE   A partition $M$ of the set of nodes that minimizes the map function $L(M)$ that is associated with $P(\vec{\delta}, t)$.
	\STATE Construct the unnormalized graph Laplacian $\mathcal{L} = (W-A)D_\delta^{-1}$ of the absorption-scaled graph with adjacency matrix $AD_\delta^{-1}$.
	\STATE Choose any Markov time $t>0$. 
	\STATE Apply the standard InfoMap algorithm with input $P(\vec{\delta},t) = e^{-t \mathcal{L}}$.
\end{algorithmic}
\end{algorithm}


\bibliography{biblio}
\bibliographystyle{plain}

\end{document}